%% file: Sargado_et_al_EnergyReleaseRate_Interface.tex
\documentclass[fleqn,3p,authoryear]{elsarticle}
\usepackage{newtxtext}
\usepackage{newtxmath}
\usepackage{amsmath}
\usepackage{esint}

\usepackage{xcolor}
\usepackage{booktabs}
\usepackage{lineno}
\usepackage{cases}
\usepackage{enumerate}

\usepackage{caption}
\usepackage[labelfont={bf,footnotesize},textfont=footnotesize]{subcaption}

\usepackage[colorlinks=true]{hyperref}

\makeatletter
\AtBeginDocument{\def\@citecolor{cyan}}
\AtBeginDocument{\def\@linkcolor{cyan}}
\makeatother

\newdefinition{remark}{Remark}

\newcommand{\strain}{\varepsilon}
\newcommand{\straintensor}{\boldsymbol{\upvarepsilon}}
\newcommand{\dee}{\mathrm{d}}
\newcommand{\contract}{\kern-2pt : \kern-2pt}

\begin{document}
\begin{frontmatter}
	\title{Approximate formulas for the energy release rate of a crack \\ perpendicular to a material interface}
	
	\author{Juan Michael Sargado\corref{cor1}}
	\ead{jmiuy@dtu.dk}
	\author{Michael Welch}
	\ead{mwelch@dtu.dk}
	\author{Mikael L\"uthje}
	\ead{mllu@dtu.dk}
	
	\cortext[cor1]{Corresponding author}
	\address{Danish Offshore Technology Centre, Technical University of Denmark, Elektrovej 375, 2800 Kongens Lyngby, Denmark}
	
	\date{}
	
	\begin{abstract}
		Rock formations are very often characterized by the presence of fractures that have grown subcritically over geological time scales and under evolving stress fields. In mechanically layered systems, such fractures can either become layer-bound or penetrate into adjacent strata. The growth of fractures in brittle materials is generally dependent on the energy release rate, however no closed form analytical solutions exist for these when a crack tip is in the proximity of a material interface. In this study, we present new empirical formulas for calculating the energy release rate at the tip of a crack perpendicular to a material interface in a symmetric 3-layer system. In these formulas, the normalized energy release rate is expressed as the product of a base term that integrates the normalized stiffness modulus over the crack length, and a correction factor that accounts for the presence of a material interface. The latter is assumed to be dependent on two quantities: the ratio of the crack length to the inner layer thickness, and the contrast in material stiffness between the inner and outer layers. The correction factors are obtained by fitting the parameters of carefully chosen expressions to a set of finite element solutions in order to yield predictions that are accurate to within one percent of the numerical results.
	\end{abstract}
	
\end{frontmatter}
\input{introduction}
\input{model_design}
\input{numerics}
\input{approximation_formulas}
\input{concluding_remarks}

\section*{Acknowledgements}
The authors kindly acknowledge the Danish Underground Consortium (TotalEnergies E\&P Denmark, Noreco \& Nordsøfonden) for granting the permission to publish this work. This research has received funding from the Danish Offshore Technology Centre (DOTC) under the AWF Improved Recovery programme.

\bibliographystyle{elsarticle-harv}
\bibliography{Reference}
\end{document}

%% file: introduction.tex
\section{Introduction}
Linear elastic fracture mechanics (LEFM) has proven to be a very successful method for analysing the growth of fractures in brittle elastic materials. It is based on the concept of energy balance -- that a fracture will propagate only if the elastic bulk energy that is unloaded per unit surface extension of the fracture (also known as the energy release rate, $G$) is equal to the energy required to break apart the solid (i.e. the material fracture toughness or critical energy release rate, $G_c$). Analytical solutions for the energy release rate of a propagating fracture were first developed by \cite{Griffith1921} for mode-I dilatant fractures in a 2D plane strain or plane stress geometry as a consequence of an applied tensile stress field. These solutions were later extended to apply to circular (penny-shaped) fractures \citep{Sack1946,Sneddon1946,SneddonElliott1946} and to mode-II shear fractures \citep{Griffith1924,Brace1960,McClintock1962}.

\begin{figure}
	\centering
	\includegraphics[width=0.6\linewidth]{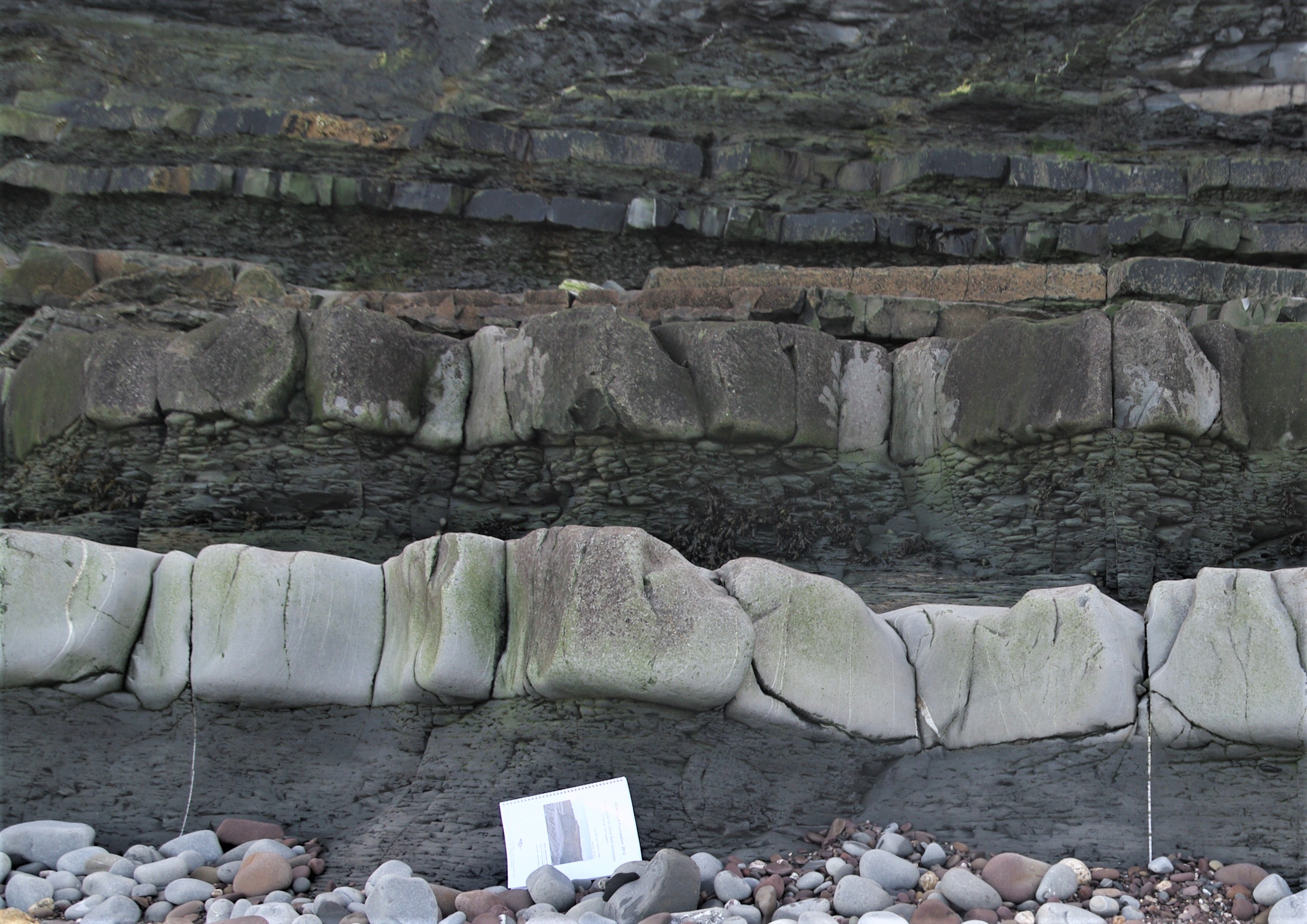}
	\caption{Mechanically layered geological system with both layer-bound and through-cutting fractures, which may be either open or sealed. Instances of the latter can be seen in the lower-left and lower-right regions of the image.}
	\label{fig:layeredFractures}
\end{figure}

LEFM was originally developed for engineering applications, but has also been adapted to model the growth of natural fractures in rocks \citep[e.g.][]{Pollard1987,Pollard1988}. In this context, it has proved useful in predicting the spatial distribution \citep[e.g.][]{Segall1984,Olsen2001,Olson2004}, geometry \citep[e.g.][]{Pollard1982,Schultz2000}, aperture \citep[e.g.][]{Bai2000,Olson2003}, and size distribution \citep[e.g.][]{Schultz2000,Welch2019} of natural fractures observed in outcrops and in the subsurface. However one major drawback with applying LEFM to geological applications is that the available analytical solutions for the energy release rate assume a homogeneous and isotropic medium, whereas real geological systems are often heterogeneous and anisotropic. A common form of heterogeneity in geological systems is mechanical layering, an example of which is shown in Figure \ref{fig:layeredFractures}.
Here the system comprises multiple horizontal (or gently inclined) layers of varying thickness, with contrasting mechanical properties (especially Young’s modulus and fracture toughness). In this situation, it is important to be able to predict whether fractures will be confined within specific mechanical layers, forming layer-bound fracture networks, or propagate past material interfaces to form through-cutting fractures. The latter are of practical importance as they can provide fluid flow pathways across otherwise impermeable layers, providing pressure communication between isolated reservoir layers, or allowing leakage of fluids such as hydrocarbons \citep[e.g.][]{Clayton1994,Gabrielsen1997} or CO2 \citep[e.g.][]{Bonto2021} from a geological trap.

Analytical solutions have been developed for linear elastic materials in which the applied stress varies along the length of the fracture, for example by \cite{England1962} and \cite{Weertman1971}. These have been successfully applied to model the propagation of fluid-driven fractures, where the fluid may be either groundwater, \cite[e.g.][]{Secor1975}, or igneous magma, \cite[e.g.][]{Pollard1976}. However these solutions still assume homogeneous mechanical properties in the rock surrounding the fracture. Heterogeneity in the mechanical properties of the rock, for example due to contrasting Young’s Modulus in different geological layers, will lead to a redistribution of the crack opening displacement along the fracture, which affects the energy release associated with the crack propagation.

For materials that are linear elastic up to the point of fracture, the application of load on a fractured specimen results in stresses that go to infinity at the crack tips. It is well known that the dominant term in such singularities is of the form $\sigma \sim r^{-p}$, where $r$ is the distance to the crack tip and $p$ is the order of the singularity. In the case of a homogeneous domain, $p = 1/2$ which leads to finite values of the energy release rate. However when the crack tip lies on an interface between two materials, the stress singularity becomes stronger or weaker depending on the stiffness contrast between the two materials. This results in the energy release rate becoming infinite if the initial layer is stiffer than the layer into which the fracture is propagating, or zero if the initial layer is more compliant \citep{Zak1963}.

Fracture propagation in heterogeneous domains has proven difficult to model analytically. However some success has been achieved by using analytical methods to reduce the problem to a set of expressions with unknown coefficients or integrals, which can then be approximated using numerical algorithms. For instance, \cite{Tamate1968} used Muskhelishvili's method to determine the stress intensity factor for a crack tip approaching an inclusion, arriving at a truncated infinite series solution whose coefficients are determined numerically. An alternative treatment of the problem was proposed by \cite{Atkinson1972}, wherein the crack is represented as as a distribution of dislocations. This leads to integral equations that must then be evaluated numerically using appropriate quadratures. The same techniques are applied in \cite{Cook1972}, \cite{Erdogan1973} and \cite{Atkinson1975} to obtain the stress intensity factor for a crack tip that is either close to impinging or has advanced a very small distance from a material interface, and later in \cite{Miller1989} to investigate crack propagation in anisotropic materials. 

More recently, hybrid approaches have been developed that utilize a combination of analytical techniques and numerical solutions for displacements and stresses, such as from finite element (FE) simulations. This include \cite{Welch2009}, which focuses on mode-II propagation of faults in layered systems, and \cite{Rijken2001}, wherein crack-bridging theory is applied to investigate the ability of thin ductile layers to arrest fractures. On the other hand, \cite{ForbesInskip2020} adopt a purely numerical approach to investigate crack arrest in layered systems, wherein fractures are represented as elliptical holes with very large aspect ratios in the FE model. This effectively blunts the cracks resulting in finite stresses at the crack tip.  However, while these techniques have had some success in modelling the growth of fracture systems in mechanically layered sections, they do not fully describe the gradual transition of the energy release rate over the full spectrum of possible locations for the crack tip along a line perpendicular to a material interface.

In this paper, we make use of finite element simulations to calculate the energy release rate for a fracture tip in different positions, before and after crossing the interface between two mechanical layers. For each fracture geometry, we run a series of models with varying contrast in the Young’s Modulus across the layer boundary. We then use the results from these models to derive general closed-form empirical expressions for the energy release rate of a propagating fracture that depend on both the crack tip distance from the interface, and the stiffness contrast between the different layers. Such formulas would allow for modeling the growth of fractures into and past layer boundaries without resorting to expensive numerical simulations, and also more accurately describe the differences in evolution over time of layer-bound versus through-cutting fractures, for example when used in combination with existing theories for subcritical crack growth.

%% file: model_design.tex
\section{Problem description}
Let us recall the basic setup of a 2-dimensional infinite elastic medium having Young's modulus $E$ and Poisson ratio $\nu$, and containing a single crack of length $2a$. It is assumed that the medium is loaded under either plane stress or plane strain conditions with a constant far-field tensile stress $\sigma_f$ (or alternatively, a far-field strain $\strain_f$) in the direction perpendicular to the crack orientation. Due to symmetry, we can consider energy quantities with respect to the half-plane. Following the work of \cite{Griffith1921}, the strain energy release due to the presence of a crack is equal to
\begin{linenomath}
\begin{equation}
	U = \frac{\pi}{2} \frac{\sigma^2}{E^\prime} a^2 = \frac{\pi}{2} E^\prime \strain_f^2 \,a^2,
\end{equation}
\end{linenomath}
with
\begin{linenomath}
\begin{equation}
	E^\prime = \left\{ \begin{array}{cl}
		E & \text{in plane stress} \\[0.5em]
		\dfrac{E}{1 - \nu^2} & \text{in plane strain}
		\end{array} \right.
\end{equation}
\end{linenomath}
being the effective Young's modulus. The strain energy release rate $G$ with respect to crack growth can be obtained by differentiating the above expression with respect to $a$:
\begin{linenomath}
\begin{equation}
	G = \frac{\partial U}{\partial a} = \pi E^\prime\strain_f^2 a,
	\label{eq:energyReleaseRate}
\end{equation}
\end{linenomath}
which implies that $G$ is linearly proportional to $a$, whereas its behavior is parabolic with respect $\strain_f$.

\begin{figure}
	\centering
	\begin{subfigure}{0.3\linewidth}
		\includegraphics[width=\textwidth]{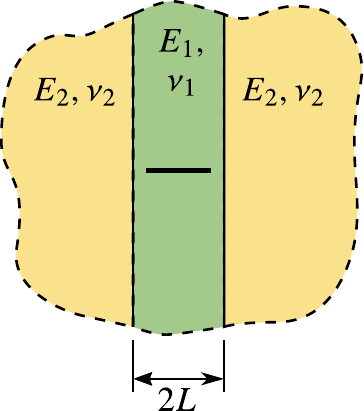}
		\caption{}
	\end{subfigure}
	\hspace{1cm}
	\begin{subfigure}{0.3\linewidth}
		\includegraphics[width=\textwidth]{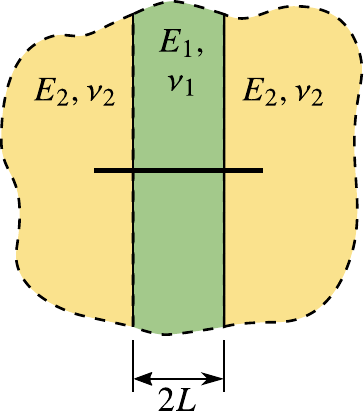}
		\caption{}
	\end{subfigure}
	\caption{Infinite heterogeneous medium containing a finite central crack. We investigate two symmetric configurations: (a) one where the crack is confined to the inner layer, and (b) one where the crack tips are situated in the outer layers. Note that the cracks are generally vertical in real life, however as the current study ignores the effect of gravity, the equivalent analysis can also be performed on a model containing horizontal fractures.}
	\label{fig:scenarios}
\end{figure}

In the current work, we analyze the case of an inner strip of width $2L$, Young's modulus $E_1$ and Poisson ratio $\nu_1$, sandwiched between two half-planes having material properties $E_2$ and $\nu_2$. We investigate two specific scenarios as illustrated in Figure \ref{fig:scenarios}: one in which a tensile crack is wholly contained within the middle layer, and one where the crack has penetrated into the outer layers.
In an actual geological formation, such fractures would most likely be vertical, however in the highly idealized setup of Figure \ref{fig:scenarios}, the geometry is rotated so that the cracks are oriented horizontally. This is possible due to the fact that we ignore the effect of gravity in the analysis. It is assumed that the crack is oriented perpendicular to the material interfaces and that the domain is loaded with a uniform tensile strain $\strain_{yy} = \strain_f$ at the far-field. No lateral constraints are applied, hence the $\sigma_{xx} = \sigma_{xy} = 0$ in the far-field stress state. We are interested primarily in the effect of material contrast as defined by the ratio of $E^\prime_1$ to $E^\prime_2$, and for simplicity we assume that $\nu_1 = \nu_2 = 0.2$. To our knowledge, no closed form solutions exist for the specific setup described above. Nevertheless, it is possible to obtain accurate numerical results by means of well-designed finite element (FE) simulations. These can in turn be utilized to construct approximate expressions for describing the dependence of $G$ on parameters such as the crack length and material contrast.

\subsection{Dimensional analysis}
As the governing equations for linear elasticity are invariant under a change of physical units, we can construct a general expression describing the dependence of the crack tip energy release rate on the various material and geometric properties by performing a dimensional analysis of the problem. This allows for the model to be expressed in terms of four dimensionless variables, i.e.
\begin{linenomath}
\begin{equation}
	F \left( \Gamma, \Upsilon, \lambda, \strain_f \right) = 0,
	\label{eq:generalModel}
\end{equation}
\end{linenomath}
where
\begin{linenomath}
\begin{equation}
	\begin{array}{ll}
	\Gamma = \dfrac{G}{E^\prime_0 L} &\text{(non-dimensional energy release rate)} \\[1.3em]
	\Upsilon = \dfrac{E^\prime}{E^\prime_0} &\text{(material contrast)} \\[1.3em]
	\lambda = \dfrac{a}{L} &\text{(normalized crack half-length)}.
	\end{array}
\end{equation}
\end{linenomath}
The quantity $E^\prime_0$ is the reference effective Young's modulus, which can be set to either $E^\prime_1$ or $E^\prime_2$. In a more general setting which permits variable Poisson ratios, $\nu_1$ and $\nu_2$ would also both need to be included as additional dimensionless variables. It can be easily verified that $G$ depends on $\varepsilon_f$ in a simple quadratic manner, even in the case where $E^\prime_1 \neq E^\prime_2$. Thus \eqref{eq:generalModel} can be simplified to
\begin{linenomath}
\begin{equation}
	\Gamma = \strain_f^2 \,f \left( \Upsilon, \lambda \right).
\end{equation}
\end{linenomath}

\subsection{General model expression} \label{sec:modelExpression}
We can observe from \eqref{eq:energyReleaseRate} that in the case of an infinite homogeneous medium containing a single crack, the energy release rate is linearly proportional to the crack length. In particular, \eqref{eq:energyReleaseRate} is equivalent to calculating the integral
\begin{linenomath}
\begin{equation}
	G  = \int_0^a \pi E^\prime \strain_f^2 \,\dee x .
\end{equation}
\end{linenomath}
We hypothesize that for a crack running through layers with varying moduli, the above expression can be recast in non-dimensional form and generalized as 
\begin{linenomath}
\begin{equation}
	\Gamma = \int_0^\lambda \pi \strain_f^2 \Upsilon \left( \xi \right) \dee\xi
	\label{eq:multilinear}
\end{equation}
\end{linenomath}
when the crack tip is sufficiently far from material interfaces ($\lambda \ll 1$ or $\lambda \gg 1$). Here $\Upsilon \left( \xi \right)$ represents the material contrast with respect to $E^\prime_0$ at the normalized distance $\xi = x/L$ along the crack. Note that in this study, we adopt the convention in which $E^\prime_0$ is chosen to be the effective Young's modulus of the layer that contains the crack tip. The dimensionless energy release rate is thus given by
\begin{linenomath}
\begin{equation}
	\Gamma = \pi \strain_f^2 \lambda
\end{equation}
\end{linenomath}
when the crack is in the inner layer, and
\begin{linenomath}
\begin{equation}
	\Gamma = \pi \strain_f^2 \left( \Upsilon + \lambda - 1 \right)
\end{equation}
\end{linenomath}
when it is in the outer layer. Here $\Upsilon = E^\prime_1 / E^\prime_2$, as it arises from the portion of \eqref{eq:multilinear} associated with the inner layer, i.e.
\begin{linenomath}
\begin{equation}
	\Gamma = \int_0^1 \pi \strain_f^2 \frac{E^\prime_1}{E^\prime_2} \dee\xi + \int_1^\lambda \pi \strain_f^2 \frac{E^\prime_2}{E^\prime_2} \dee\xi.
\end{equation}
\end{linenomath}

On the other hand when the aforementioned remoteness condition is not fulfilled, we calculate the energy release rate as a combination of two effects: a base contribution given by \eqref{eq:multilinear}, together with a further scaling to account for proximity to the interface. Thus when the crack tip is within the inner layer, the relevant expression is
\begin{linenomath}
\begin{equation}
	\Gamma = \pi\strain_f^2 \lambda f_1 \left( \Upsilon, \lambda \right)
	\label{eq:normalizedEnergyReleaseRate_1}
\end{equation}
\end{linenomath}
in which $\Upsilon = E^\prime_2 / E^\prime_1$, whereas when the crack tip resides within the outer layer, the energy release rate is calculated as
\begin{linenomath}
\begin{equation}
	\Gamma = \pi\strain_f^2 \left( \Upsilon + \lambda - 1  \right) f_2 \left( \Upsilon, \lambda \right),
	\label{eq:normalizedEnergyReleaseRate_2}
\end{equation}
\end{linenomath}
where now $\Upsilon = E^\prime_1 / E^\prime_2$ as explained earlier. The last two results above imply that $f_1 \left( \Upsilon, \lambda \right)$ should tend toward 1 as $\lambda \rightarrow 0$, and $f_2 \left( \Upsilon, \lambda \right)$ should tend toward 1 as $\lambda \rightarrow \infty$, in order to recover \eqref{eq:multilinear} when either $\lambda \ll 1$ or $\lambda \gg 1$.

%% file: numerics.tex
\section{Numerical simulations}
\subsection{Discrete problem setup and FE solution}
\begin{figure*}
	\centering
	\begin{subfigure}{0.3\linewidth}
		\includegraphics[width=\textwidth]{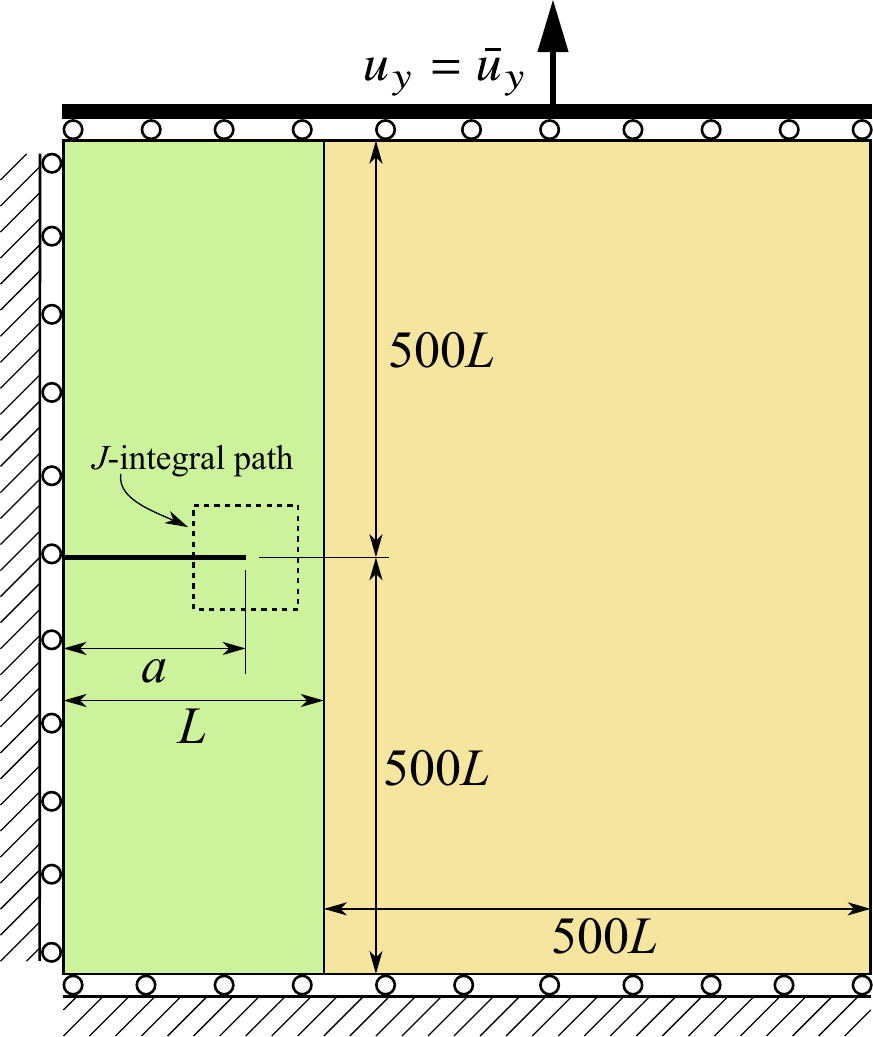}
		\caption{}
	\end{subfigure}
	\hspace{10mm}
	\begin{subfigure}{0.3\linewidth}
		\includegraphics[width=\textwidth]{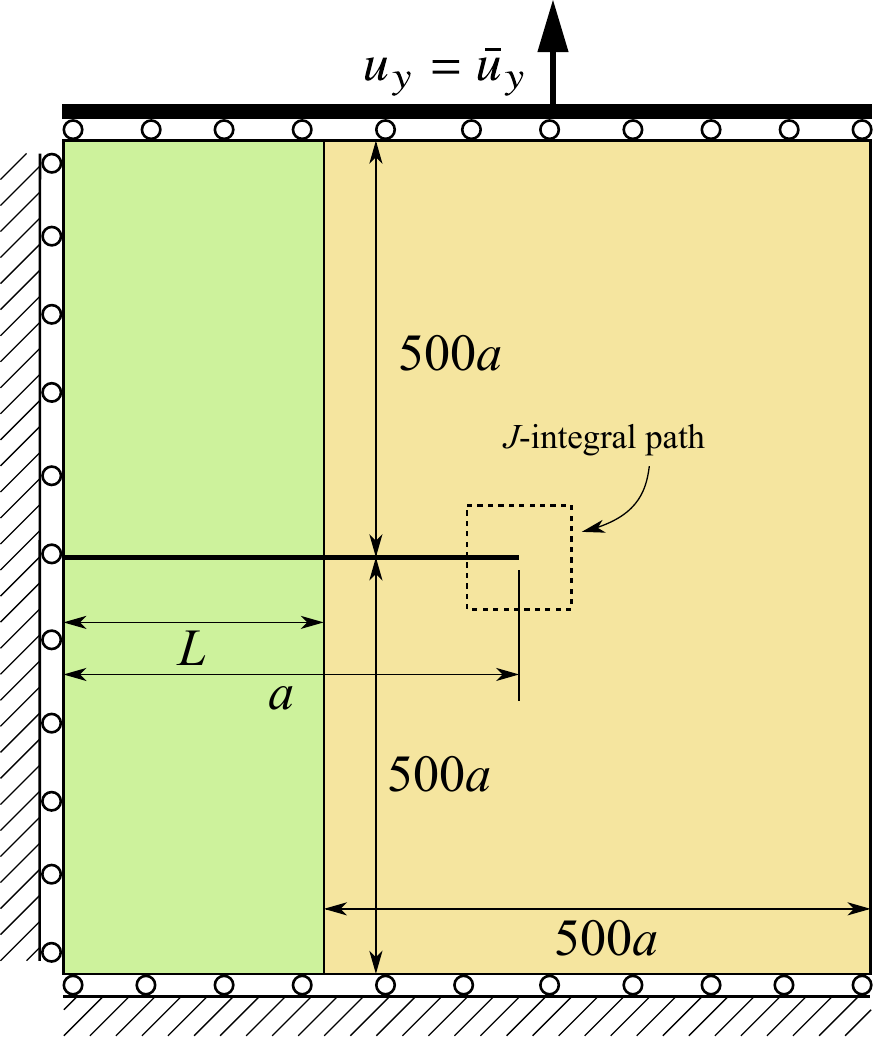}
		\caption{}
	\end{subfigure}
	\caption{Computational domains with applied boundary conditions for (a) $a < L$, and (b) $a > L$. The prescribed upward displacement $\bar{u}_y$ is chosen such that a uniform strain $\strain_{yy} = 0.1$ is realized at the top and bottom boundaries. Thus, $\bar{u}_y = 100 L$ when $a < L$, while $\bar{u}_y = 100 a$ when $a > L$.}
	\label{fig:compDomains}
\end{figure*}

\begin{figure}
	\centering
	\includegraphics[width=\linewidth]{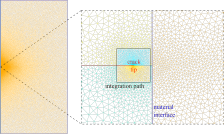}
	\caption{Meshing details for $a = 0.999L$. The full computational domain is shown at the left, together with a closeup of the crack-tip region. For clarity, the crack is highlighted in red, the material interface in blue, and the integration path for the $J$-integral in black. For all meshes, the closed path used to calculate the $J$-integral is constructed to lie fully within the material layer containing the crack tip, as illustrated in Figure \ref{fig:compDomains}.}
	\label{fig:meshDetail}
\end{figure}

To determine the energy release rates corresponding with different crack lengths and Young's moduli in the symmetric 3-layered medium described in the previous section, we resort to a numerical approach by means of finite element (FE) simulations. Due to the inherent symmetry of the problem, only half of the actual domain needs to be modeled. However as the original problem involves an infinite medium, the overall dimensions of the computational domain need to be taken sufficiently large with respect to the crack length in order to properly model the far-field state of stress. We consider two types of computational domains, which differ on whether or not the crack has gone past the material interface. These are illustrated in Figure \ref{fig:compDomains}.
When specifying the various geometrical entities, we assume that the lower left corner of the overall domain is located at $\left( 0,0 \right)$.
Note that whereas in a homogeneous medium it is possible to obtain a sufficiently accurate approximation of the far-field stress state by truncating the computational domain to dimensions of around an order of magnitude larger than the crack length, for the specific problem considered in the present work we have found it necessary to extend the dimensions of computational domains to between two and three orders of magnitude larger than the crack length. This requires a careful approach with regard to discretization, since constructing meshes with uniformly sized elements would result in a prohibitively large number of unknowns to be solved. Instead, we utilize a combination of structured and unstructured triangulation strategies together with non-uniform mesh refinement. An example of this is shown in Figure \ref{fig:meshDetail}, wherein the characteristic size of elements is reduced at regions of expected high stress gradients.

\begin{figure}
	\centering
	\begin{subfigure}{0.3\linewidth}
		\includegraphics[width=\textwidth]{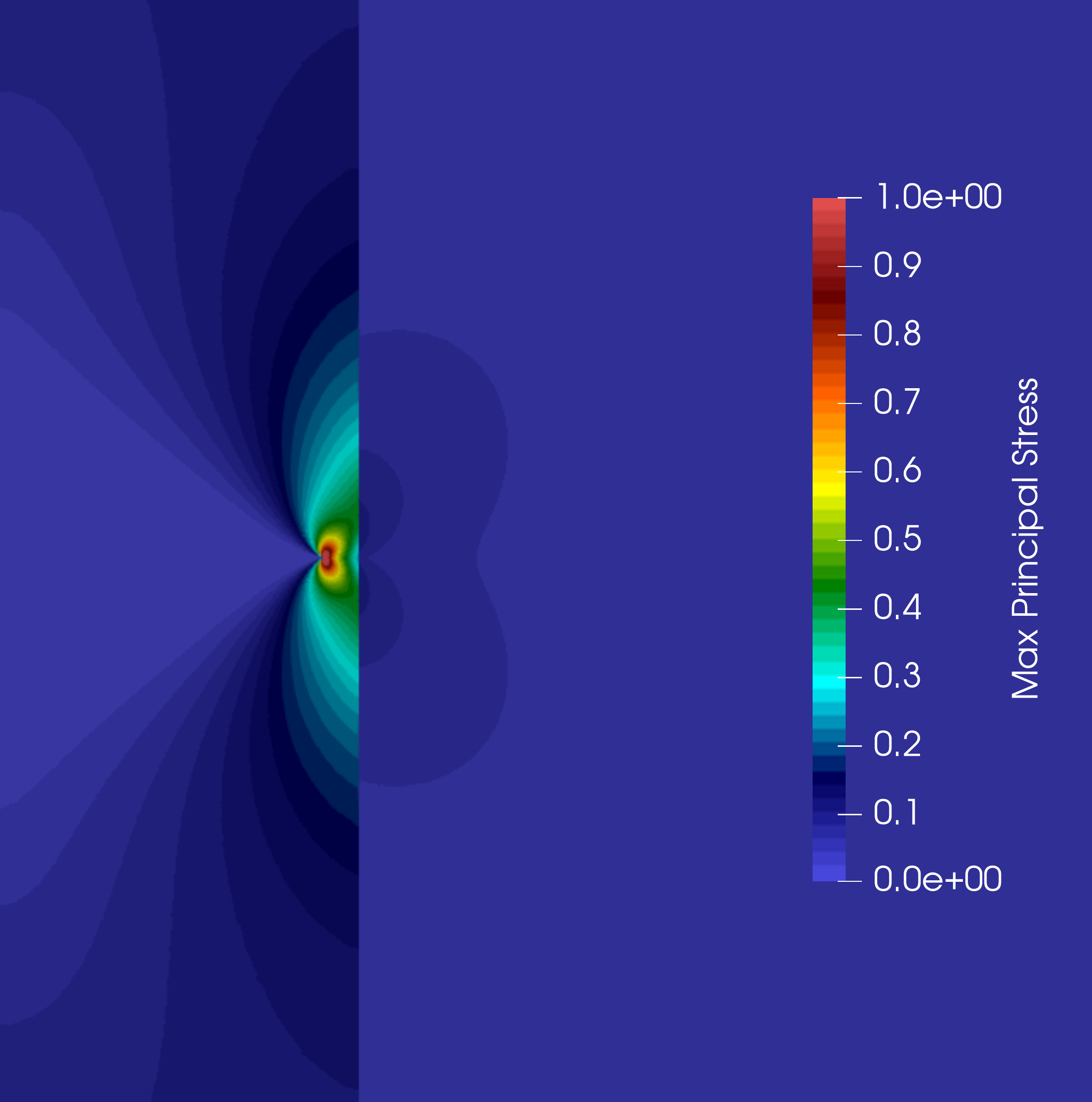}
		\caption{}
	\end{subfigure} \hspace{1cm}
	\begin{subfigure}{0.3\linewidth}
		\includegraphics[width=\textwidth]{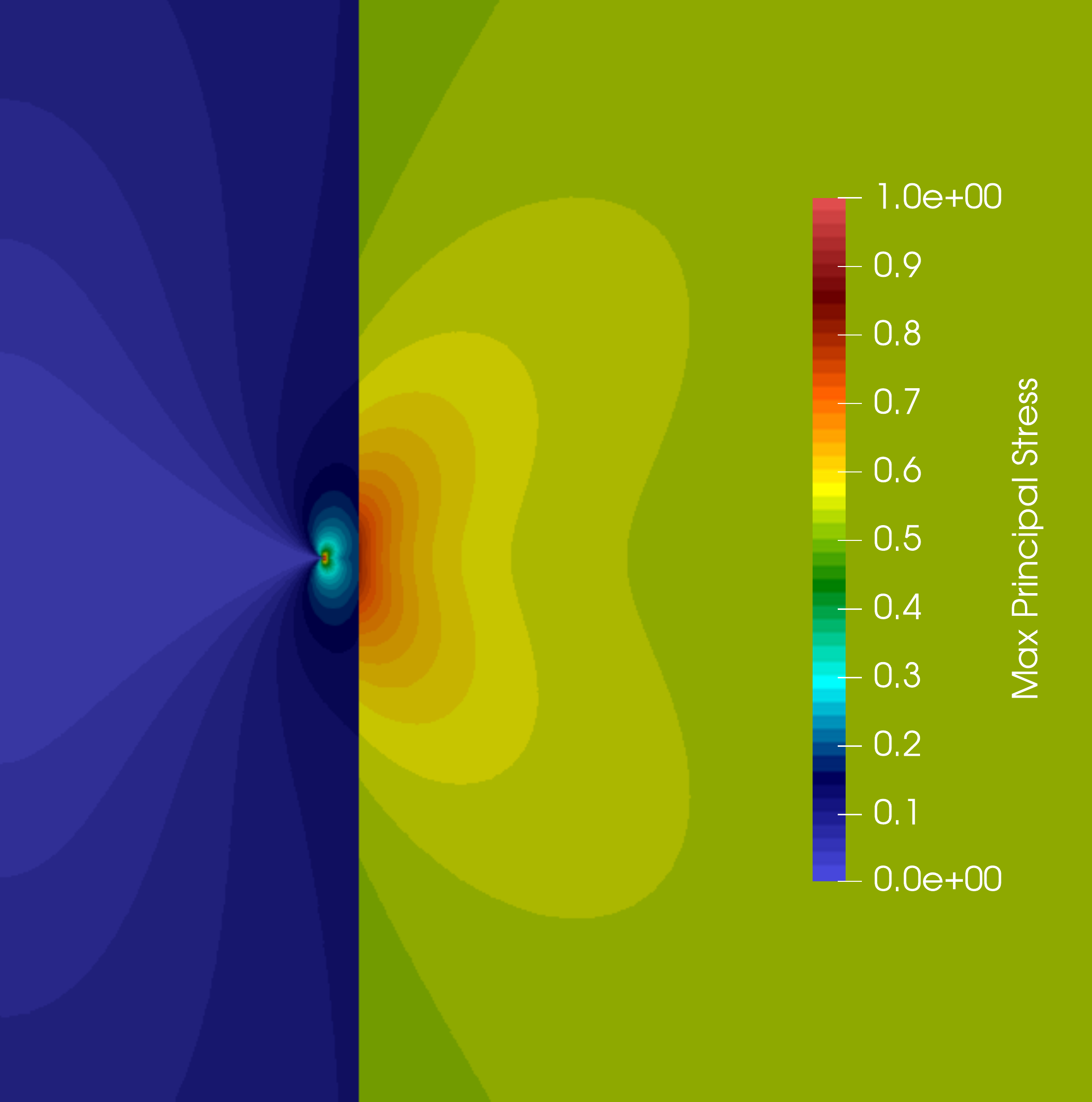}
		\caption{}
	\end{subfigure} \\[1em]
	\begin{subfigure}{0.3\linewidth}
		\includegraphics[width=\textwidth]{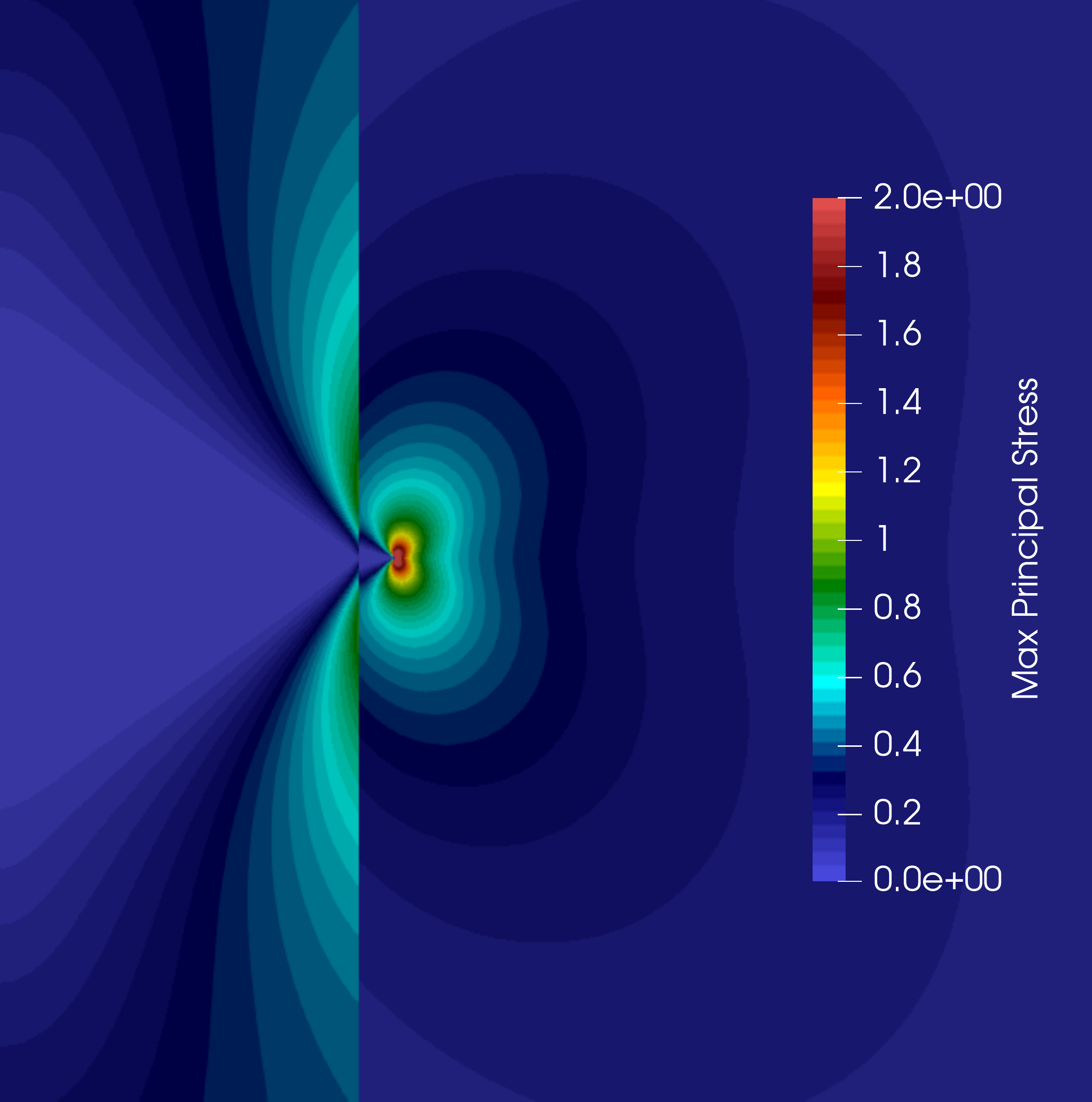}
		\caption{}
	\end{subfigure} \hspace{1cm}
	\begin{subfigure}{0.3\linewidth}
		\includegraphics[width=\textwidth]{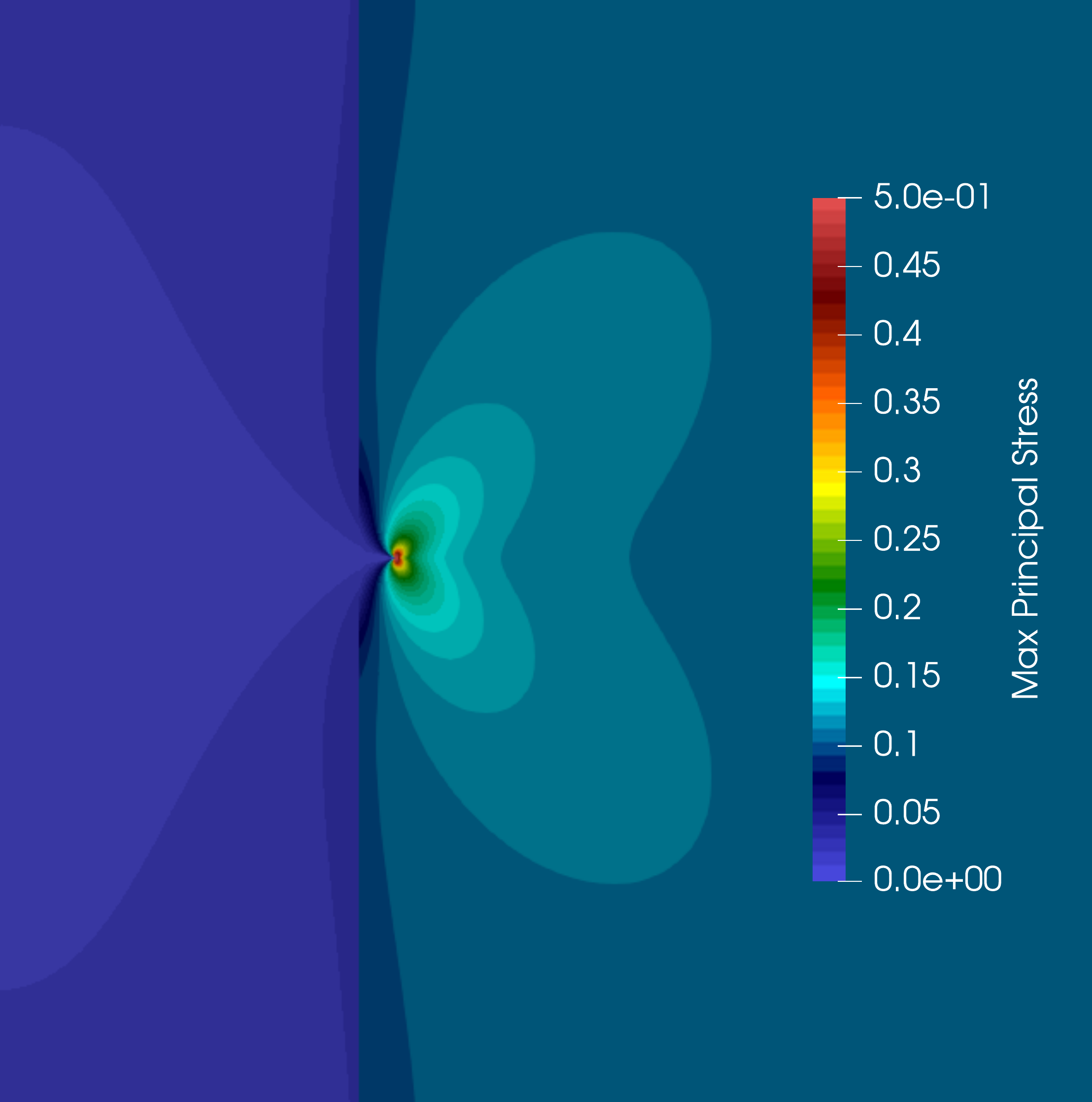}
		\caption{}
	\end{subfigure}
	\caption{Variation in maximum principal stress in the crack tip vicinity for some combinations of the normalized crack length and material stiffness contrast: (a) $\Upsilon = 5$, $\lambda = 0.9$; (b) $\Upsilon = 0.2$, $\lambda = 0.9$; (c) $\Upsilon = 5$, $\lambda = 1.1$; (d) $\Upsilon = 0.2$, $\lambda = 1.1$.}
	\label{fig:stress_variation}
\end{figure}

The primary unknown to be solved is the displacement field $\mathbf{u} = \{ u, v \}^\mathrm{T}$, which is approximated as a piecewise continuous function over the mesh using $P_2$ Lagrange elements (6-node triangles).  The governing equations consist of linear momentum conservation in 2-d, given in weak form as
\begin{linenomath}
\begin{equation}
	\int_\Omega \delta\straintensor\contract\mathbb{C} \contract\straintensor \,\dee\Omega = \int_\Omega \delta \mathbf{u} \cdot \mathbf{b} \,\dee\Omega + \int_{\partial\Omega_N} \delta \mathbf{u} \cdot \mathbf{t} \,\dee S .
	\label{eq:weakForm}
\end{equation}
\end{linenomath}
In the above expression, $\straintensor$ is the symmetric small strain tensor, $\delta\mathbf{u}$ and $\delta\straintensor$ respectively denote the variations of the displacement and strain fields, $\mathbf{b}$ is the body force, and $\mathbf{t}$ is the prescribed surface traction acting on the Neumann boundary $\partial\Omega_N$. The discrete approximation of the weak form is obtained by first approximating the displacement field and its variation over each element as
\begin{linenomath}
\begin{equation}
	\mathbf{u} = \left[ \mathbf{N}^e \right] \left\{ \mathbf{u}^e \right\}, \qquad \delta \mathbf{u} = \left[ \mathbf{N}^e \right] \left\{ \delta \mathbf{u}^e \right\}.
\end{equation}
\end{linenomath}
The nodal displacements are ordered as $\{ \mathbf{u}^e \} = \{ u_1, v_1, \ldots, u_6, v_6 \}^\mathrm{T}$, and
\begin{linenomath}
\begin{equation}
	\left[ \mathbf{N}^e \right] = \left[ \begin{array}{ccccc}
		\psi^e_1 & 0 & \ldots & \psi^e_6 & 0 \\[5pt]
		0 & \psi^e_1 & \ldots & 0 & \psi^e_6
	\end{array} \right]
\end{equation}
\end{linenomath}
is the approximation matrix that interpolates the nodal values to any location within $\Omega^e$ by means of the element shape functions $\psi_I^e$. The strain tensor is expressed in Voigt form as $\{ \straintensor \} = \{ \varepsilon_{xx}, \varepsilon_{yy}, \gamma_{xy} \}$, and both the strain and its variation are calculated from nodal values of $\mathbf{u}$ and $\delta\mathbf{u}$ as
\begin{linenomath}
\begin{equation}
	\left\{ \straintensor \right\} = \left[ \mathbf{B}^e \right] \left\{ \mathbf{u}^e \right\}, \qquad
	\left\{ \delta \straintensor \right\} = \left[ \mathbf{B}^e \right] \left\{ \delta \mathbf{u}^e \right\},
\end{equation}
\end{linenomath} 
in which $\left[ \mathbf{B}^e \right]$ is the gradient matrix given by
\begin{linenomath}
\begin{equation}
	\left[ \mathbf{B}^e \right] = \left[ \begin{array}{ccccc}
		\dfrac{\partial\psi^e_1}{\partial x} & 0 & \ldots & \dfrac{\partial\psi^e_6}{\partial x} & 0 \\[5pt]
		0 & \dfrac{\partial\psi^e_1}{\partial y} & \ldots & 0 & \dfrac{\partial\psi^e_6}{\partial y} \\[8pt]
		\dfrac{\partial\psi^e_1}{\partial y} & \dfrac{\partial\psi^e_1}{\partial x} & \ldots & \dfrac{\partial\psi^e_6}{\partial y} & \dfrac{\partial\psi^e_6}{\partial x}
	\end{array} \right].
\end{equation}
\end{linenomath}
The weak form should hold for arbitrary values of $\delta\mathbf{u}$, thus substitution of the above expressions into \eqref{eq:weakForm} leads to the sparse linear system
\begin{linenomath}
\begin{equation}
	\left[ \mathbf{K} \right] \left\{ \mathbf{U} \right\} = \left\{ \mathbf{F} \right\},
	\label{eq:linearSystem}
\end{equation}
\end{linenomath}
where the global coefficient matrix $\left[ \mathbf{K} \right]$ and right hand side $\left\{ \mathbf{F} \right\}$ are obtained by assembling local contributions $\left[ \mathbf{K}^e \right]$ and $\left\{ \mathbf{F}^e \right\}$ from each element. The latter are calculated as
\begin{linenomath}
\begin{align}
	\left[ \mathbf{K}^e \right] &= \int_{\Omega^e} \left[ \mathbf{B}^e \right]^\mathrm{T} \left[ \mathbf{C} \right] \left[ \mathbf{B}^e \right] \dee\Omega \\
	\left\{ \mathbf{F}^e \right\} &= \int_{\Omega^e} \left[ \mathbf{N^e} \right]^\mathrm{T} \left\{ \mathbf{b} \right\} \dee\Omega + \int_{\partial\Omega^e} \left[ \mathbf{N^e} \right]^\mathrm{T} \left\{ \mathbf{t} \right\} \dee S,
\end{align}
\end{linenomath}
where $\left[ \mathbf{C} \right]$ is the Voigt form of the elasticity tensor assuming plane strain conditions: 
\begin{linenomath}
\begin{equation}
	\left[ \mathbf{C} \right] = \frac{E}{\left( 1 + \nu \right) \left( 1 - 2\nu \right)} \left[ \begin{array}{ccc}
		1 - \nu & \nu & 0 \\[5pt]
		\nu & 1 - \nu & 0 \\[5pt]
		0 & 0 & \dfrac{1 - 2\nu}{2}
	\end{array} \right].
\end{equation}
\end{linenomath}
For linear elastic materials, it is common knowledge that stresses are singular at crack tips and thus cannot be properly captured by standard approximation techniques. A common practice is to use special quarter-point elements originally developed by \cite{Barsoum1977}  at the crack tip region in order to recover optimal convergence rates. However with the computing power and memory available in current machines, accurate solutions can also be achieved by utilizing standard elements, provided that the mesh is sufficiently refined at the regions where stresses are singular. This is the strategy adopted in the current work, where the characteristic element size $h^e$ at the crack tip is set to between $10^{-5} a$ and $10^{-4} a$ depending on the actual length of the crack. Figure \ref{fig:stress_variation} shows the variation in maximum principal stress for different combinations of crack lengths and material stiffness contrast.
We can see that when the crack tip is in close proximity to the material interface and lies within a more compliant layer, high stress gradients occur not only in the immediate crack tip vicinity, but also in the opposite layer across the interface. This must be accounted for when designing the mesh refinement so as to produce sufficiently accurate results.

After solution of the unknown nodal displacements, we determine the mode-I energy release rate at the crack tip by evaluating the $J$-integral of \cite{Cherepanov1967} and \cite{Rice1968}. For a straight crack parallel to axis $\xi$, the $J$-integral is calculated as
\begin{linenomath}
\begin{equation}
	J = \oint_C \left( \Psi n_\xi - \mathbf{t} \cdot \frac{\partial \mathbf{u}}{\partial \xi} \right) \dee S
	\label{eq:Jintegral}
\end{equation}
\end{linenomath}
where $C$ is a closed contour that encloses the crack tip, $\Psi$ denotes the strain energy density, $n_\xi$ is the $\xi$-component of the unit normal to $\dee S$,  $\mathbf{t}$ denotes the surface traction, and $\partial \mathbf{u} \, / \partial \xi$ is the displacement gradient component along $\xi$. In practice, the computational domain is discretized in such a way that $C$ coincides with edges of the triangular elements. As quantities derived from the displacement gradient are discontinuous over element edges, the required values of stress and strain components along the integration path are obtained by averaging the values associated with the two elements sharing a particular edge.

\subsection{Summary of numerical results}
The finite element simulations are performed using the open-source code BROOMStyx \citep{Sargado2019}, which runs in parallel on multi-core processors via OpenMP.  Specific values are chosen for the geometric parameters seen in Figure \ref{fig:compDomains}, as well as the material properties of both the inner and outer layers. For convenience, we group the results into four categories based on the ratio of geometric and material paramaters, i.e.
\begin{itemize}
	\setlength{\itemsep}{0pt}
	\item Case I-a: $a < L$ and $E_1 < E_2$
	\item Case I-b: $a < L$ and $E_1 > E_2$
	\item Case II-a: $a > L$ and $E_1 < E_2$
	\item Case II-b: $a > L$ and $E_1 > E_2$
\end{itemize}
For cases I-a and I-b, we let $L = 10$, $E_1 = 1$, $\nu_1 = \nu_2 = 0.2$, and run simulations for different values of $a$, and $E_2$. Specifically, $a \in $ \{1, 2, 3, 4, 5, 6, 7, 8, 9, 9.5, 9.9, 9.99\} while $E_2 \in $ \{1/100, 1/14, 1/12, 1/10, 1/8, 1/6, 1/4, 1/2, 2/3, 1, 5/4, 5/3, 5/2, 10, 20, 40\}. Additionally, there are two limit cases to consider: one in which $E_2 \rightarrow 0$, and another where $E_2 \rightarrow \infty$. Both cases are modeled by removing the outer layer and imposing appropriate conditions on the material interface, which acts as a boundary. For $E_2 \rightarrow 0$, the problem reduces to a homogeneous center-cracked specimen of width $2L$, with the original material interface acting as free boundary. On the other hand, for $E_2 \rightarrow \infty$, the interface must move horizontally as a rigid body, while vertical displacements are linear with respect to the $y$-coordinate. That is, $u_x \left( L,y \right) = A$ where $A$ is some constant to be solved, and $u_y \left( L,y \right) = 0.1 y$ so that $\strain_{yy} = 0.1$. These conditions are inferred by considering the alternate problem wherein $E_2$ is re-scaled to unity so that $E_1 \rightarrow 0$. In this case, the inner layer with the crack can be ignored and the problem reduces to the trivial case of a homogeneous domain (without any crack) subject to a uniform state of strain with the shear component $\gamma_{xy}$ equal to zero. This implies that for $x > L$, $u_x \left( x,y \right) = Ax$ and $u_y \left( x,y \right) = By$, where $A, B \in \mathbb{R}$.

\begin{figure}
	\centering
	\includegraphics[width=0.5\linewidth]{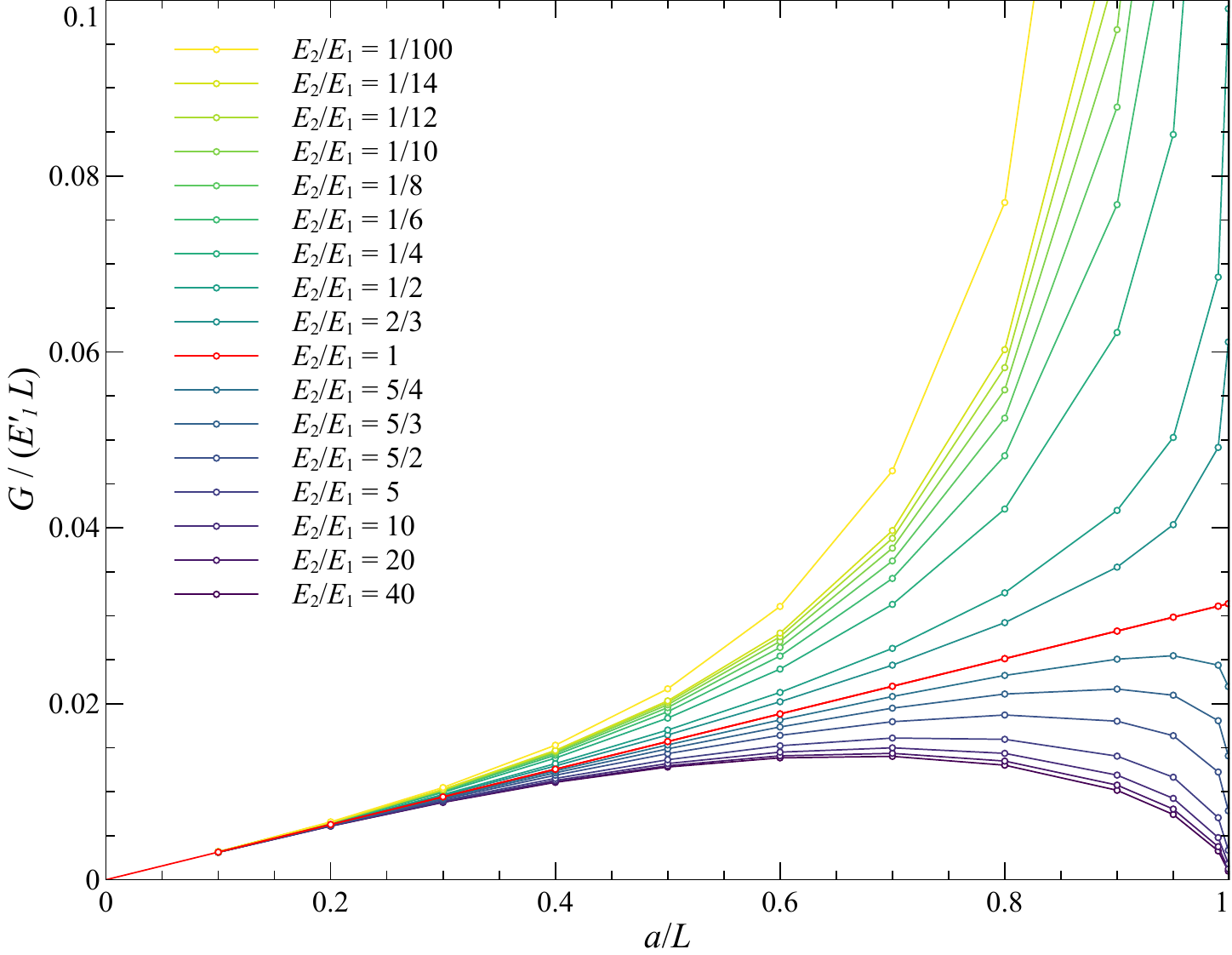}
	\caption{Calculated energy release rates for $a/L \leq 1$, normalized with respect to $E^\prime_1$.}
	\label{fig:nondimResults_1}
\end{figure}

In the FE simulations, rigid body motion is imposed by means of master/slave-type constraints on the relevant degrees of freedom at the affected boundary. For generality, the results are presented in terms of dimensionless quantities $\Gamma$, $\Upsilon$ and $\lambda$ in Figure \ref{fig:nondimResults_1}.
Following the discussion in Section \ref{sec:modelExpression}, the dimensionless energy release rate is normalized with respect to $E_1^\prime$ since the crack tip is located within the inner layer. We can see that for $E_1 < E_2$, the energy release rate converges to zero as the crack tip gets closer to the material interface, while the same goes to infinity when $E_1 > E_2$. The observed trends thus verify the analytical result of \cite{Zak1963} pertaining to the energy release rate at a crack tip lying on a material interface.

Meanwhile for cases II-a and II-b, the simulations are carried out with $L = 10$ and $E_2 = 1$, with $\nu = 0.2$ for both inner and outer layers. We run simulations corresponding to different values of $a$ and $E_1$. Here we choose $a \in$ \{10.01, 10.025, 10.05, 10.1, 10.2, 10.5, 11, 12, 13, 14, 15, 20, 30, 50, 80, 110, 160, 210, 310\}, and $E_1 \in$ \{0.025, 0.05, 0.1, 0.2, 0.4, 0.6, 0.8, 1, 1.5, 2, 4, 6, 8, 10, 12, 14, 16, 18, 20, 50, 100\}. In the limit where $E_1 \rightarrow 0$, the problem reduces to the special case of the double edge notch test specimen having infinite width (see for example, \cite{Tada1973}). Note however that for $a > L$, no limit case corresponding to $E_1 \rightarrow \infty$ exists for which a uniform far-field state of strain with $\strain_{yy} > 0$ can be achieved. This becomes clear when we adopt the earlier trick of re-scaling the elastic moduli such that $E_1 = 1$ and $E_2 \rightarrow 0$. Here the problem reduces to one involving two unconnected subdomains that are incapable of generating the needed boundary reactions to equilibrate the surface tractions arising from a nonzero vertical stress component at the far-field boundary.

\begin{figure*}
	\centering
	\begin{subfigure}{0.47\linewidth}
		\includegraphics[width=\textwidth]{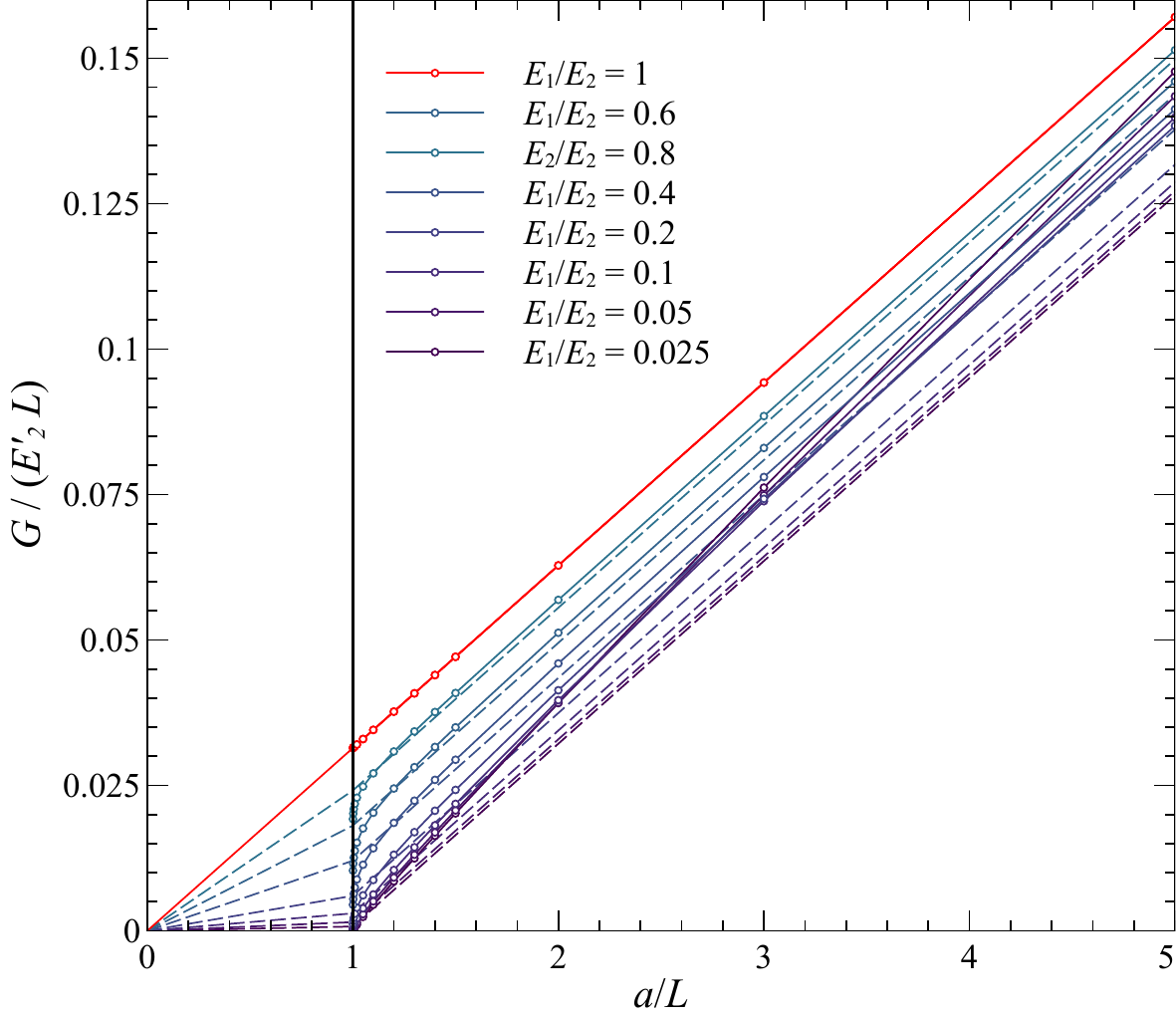}
		\caption{}
		\label{fig:nondimResults_2A}
	\end{subfigure} \hspace{5mm}
	\begin{subfigure}{0.47\linewidth}
		\includegraphics[width=\textwidth]{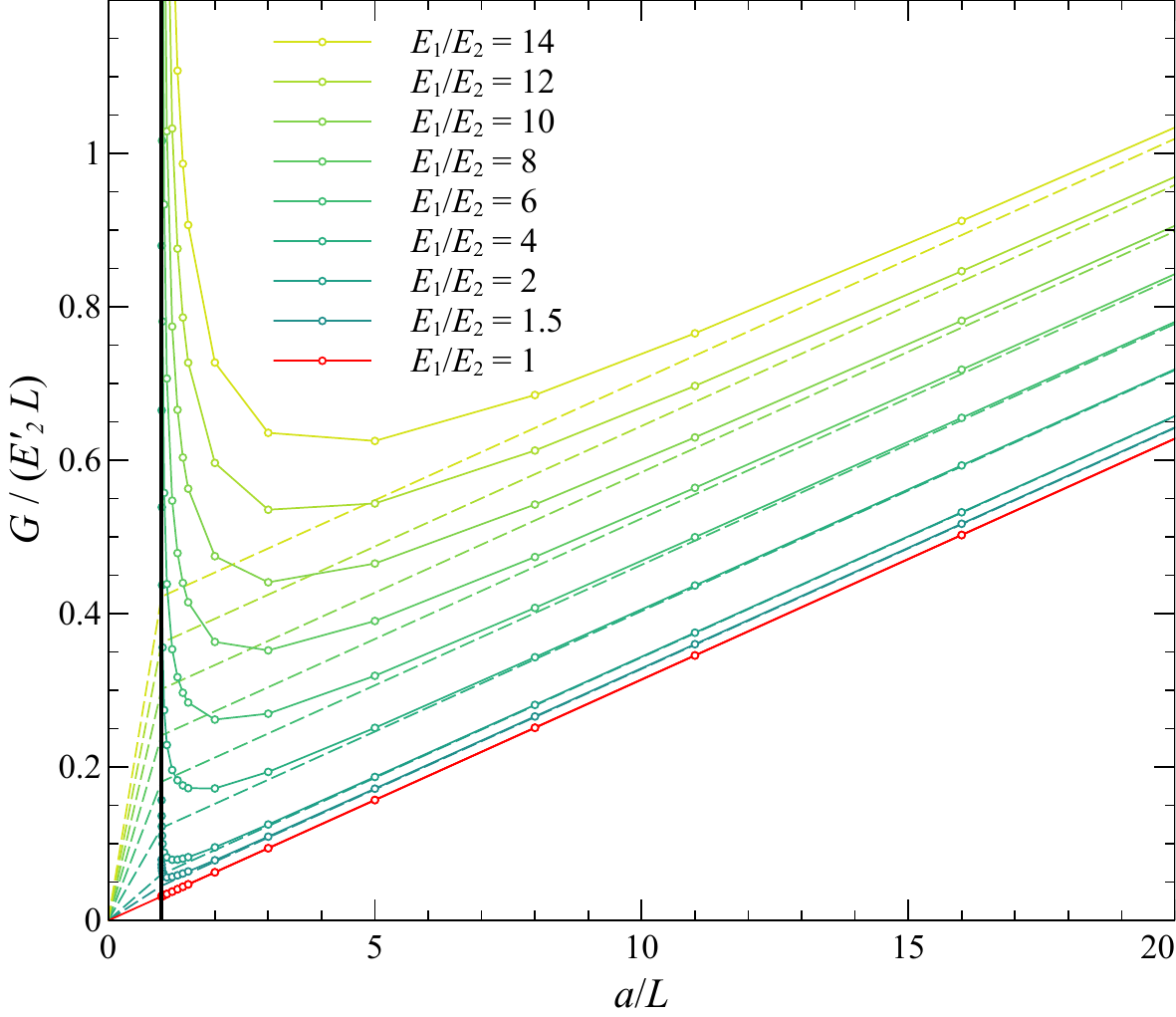}
		\caption{}
		\label{fig:nondimResults_2B}
	\end{subfigure}
	\caption{Calculated energy release rates normalized with respect to $E^\prime_2$, for $a/L > 1$ and (a) $E^\prime_1/E^\prime_2 < 1$, (b) $E^\prime_1/E^\prime_2 > 1$. The dashed lines represent the curves obtained by evaluating \eqref{eq:normalizedEnergyReleaseRate_1} and \eqref{eq:normalizedEnergyReleaseRate_2} with the interface correction factors $f_1$ and $f_2$ set to unity.}
	\label{fig:nondimResults_2}
\end{figure*}

Similar to the earlier cases, we plot the numerical results using dimensionless quantities, where it should be noted that the normalization of $\Gamma$ is now done with respect to $E_2^\prime$, since the crack tip resides in the outer layer. These results are shown in Figure \ref{fig:nondimResults_2}.
For clarity, we construct separate plots for the data on each side of $\Upsilon = 1$, as the dependence $\Gamma$ on $\lambda$ and $\Upsilon$ is markedly different going from case II-a to II-b. In particular when $\Upsilon < 1$, the calculated values for $\Gamma$ overshoot the lines representing $\pi \strain_f^2 \left( \Upsilon + \lambda - 1 \right)$, indicating that $f_2 \left( \Upsilon < 1, \lambda \right)$ is not bounded by unity from above. This does not necessarily invalidate the applicability of \eqref{eq:normalizedEnergyReleaseRate_2} to case II-a, however it does imply non-monotonicity of the interface correction factor, in contrast to what is observed in cases I-a, I-b and II-b.

%% file: approximation_formulas.tex
\section{Closed form approximation of numerical results}
\begin{figure*}
	\centering
	\includegraphics[width=0.7\textwidth]{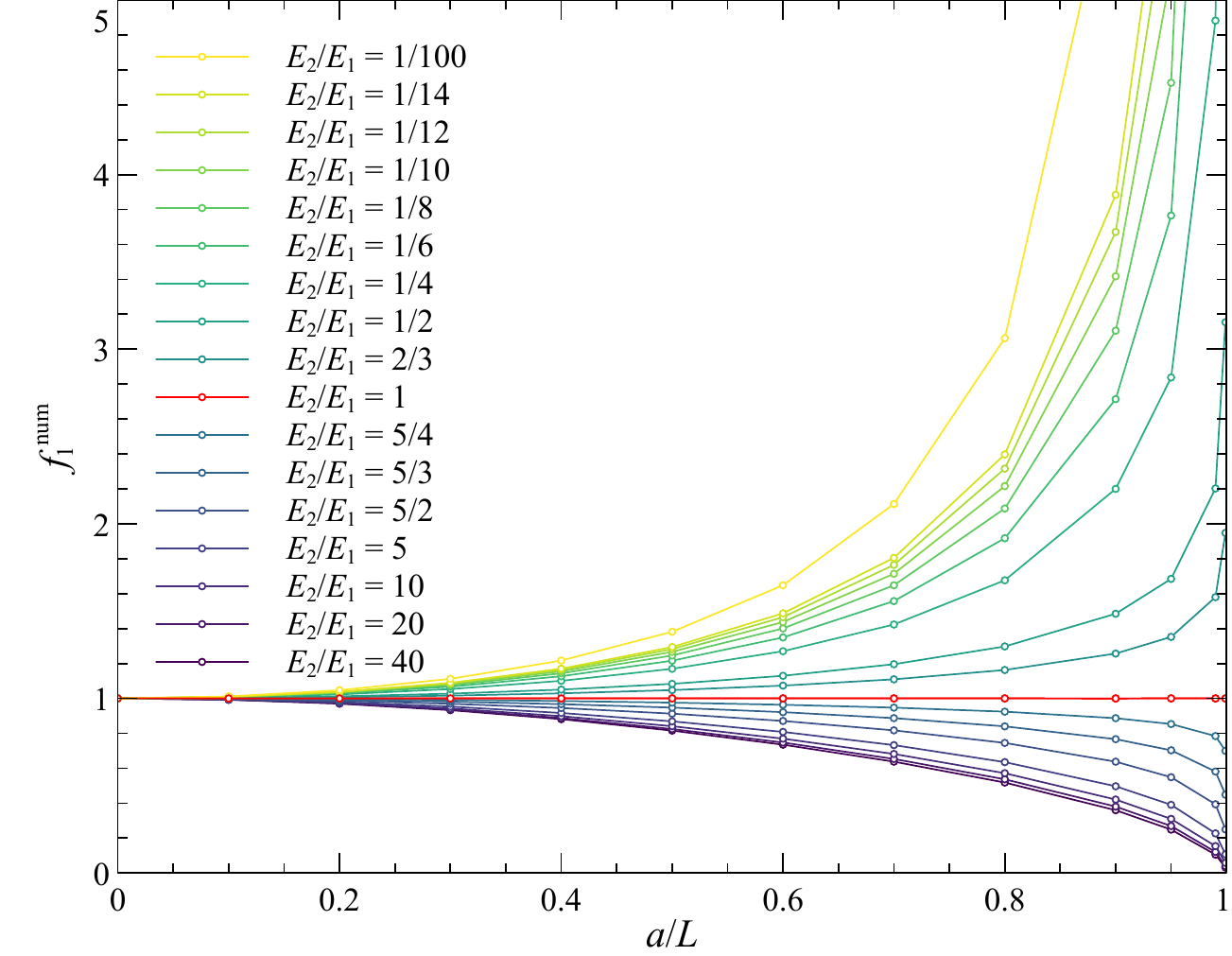}
	\caption{Interface correction factor $f_1^\text{num}$ calculated from FE results for $a < L$.}
	\label{fig:fnum_1}
\end{figure*}

\begin{figure*}
	\centering
	\includegraphics[width=0.7\textwidth]{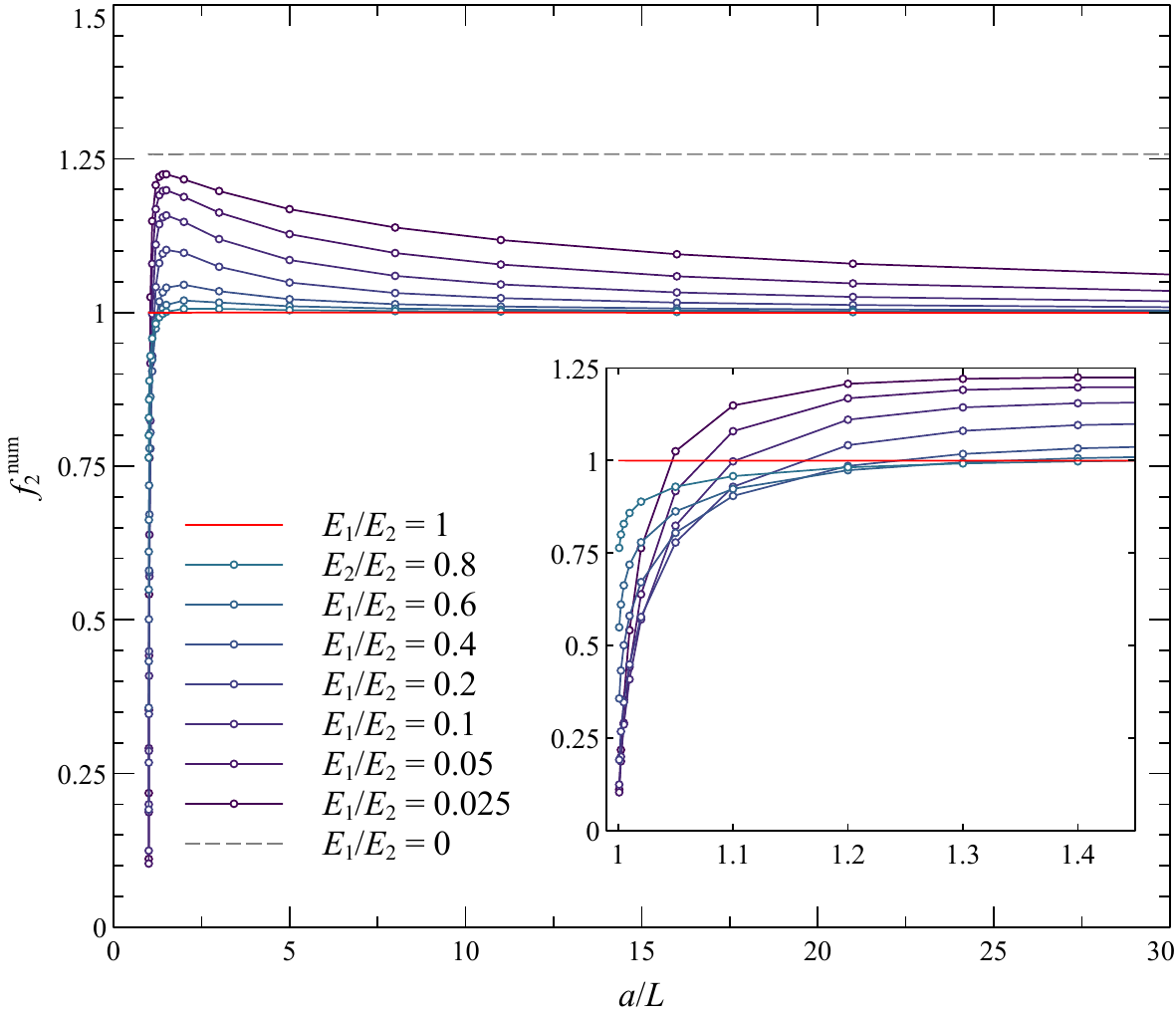}
	\caption{Interface correction factor $f_2^\text{num}$ calculated from FE results for $a > L$ and $E_1 < E_2$. Inset shows behavior of $f_2^\text{num}$ for values of $a/L$ close to unity.}
	\label{fig:fnum_2A}
\end{figure*}

\begin{figure*}
	\centering
	\includegraphics[width=0.69\textwidth]{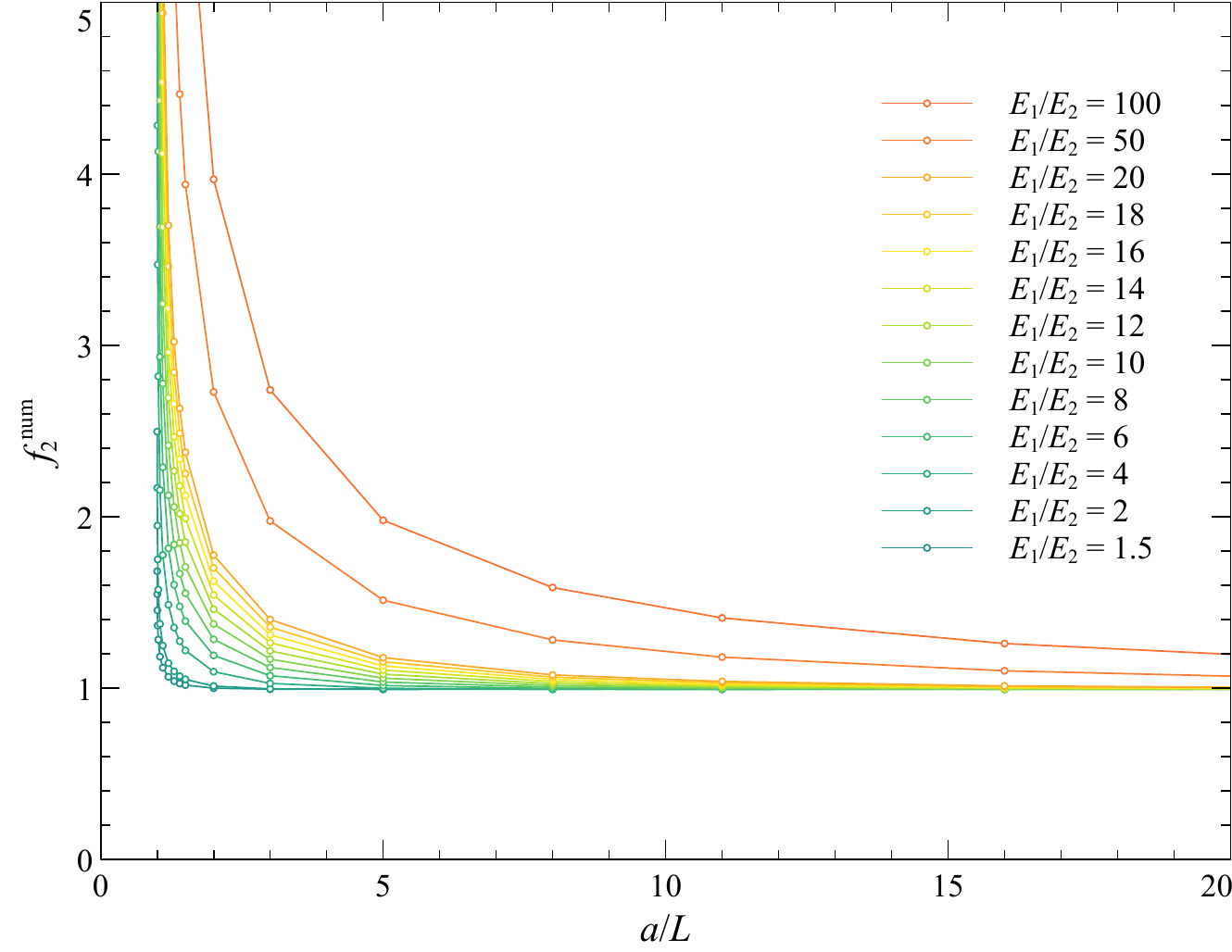}
	\caption{Interface correction factor $f_2^\text{num}$ calculated from FE results for $a > L$ and $E_1 > E_2$.}
	\label{fig:fnum_2B}
\end{figure*}

\begin{figure*}
	\centering
	\includegraphics[width=0.69\textwidth]{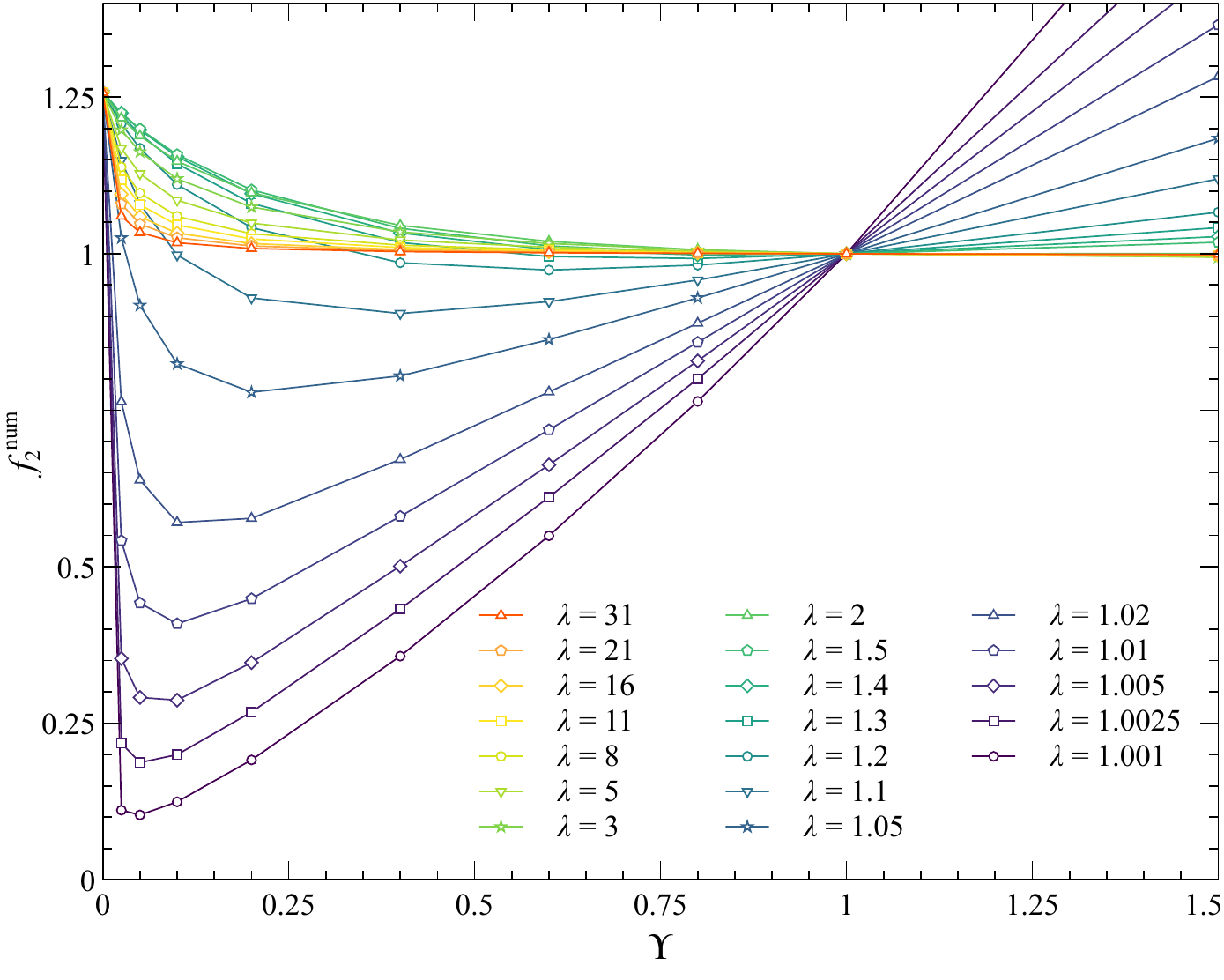}
	\caption{Dependence of the interface correction factor on the material contrast for $a > L$. Note that $f_2^\text{num}$ is generally non-monotone with respect to $\Upsilon$ when $\Upsilon < 1$, but becomes a strictly increasing function with respect to $\Upsilon$ when $\Upsilon > 1$.}
	\label{fig:nonmonotonicity} 
\end{figure*}

In this section, we construct closed form approximations for the energy release rates pertaining to the setup illustrated in Figure \ref{fig:scenarios}  based on the general model proposed in Section \ref{sec:modelExpression} and utilizing the numerical results reported in the previous section. To accomplish this, we calculate the numerical approximation of the interface correction factors as
\begin{linenomath}
\begin{equation}
	f_1^\text{num} = \frac{\Gamma^\text{\,num}}{\pi \strain_f^2 \lambda}
\end{equation}
\end{linenomath}
for cases I-a and I-b, and
\begin{linenomath}
	\begin{equation}
		f_1^\text{num} = \frac{\Gamma^\text{\,num}}{\pi \strain_f^2 \left( \Upsilon + \lambda - 1 \right)}
	\end{equation}
\end{linenomath}
for cases II-a and II-b. The resulting values are plotted in Figures \ref{fig:fnum_1}--\ref{fig:fnum_2B}.
As mentioned previously, the interface correction factor appears to exhibit monotonicity in all cases except in II-a, where its behavior is significantly more complex. Notably, $f_2$ converges to 0 at $\lambda = 1$, achieves a peak value greater than 1 at some $\lambda$ between around 1.4 and 2, then finally decreases asymptotically to 1 as $\lambda \rightarrow \infty$. The dashed gray line in Figure \ref{fig:fnum_2A} represents the correction factor associated with limiting case where $E_1 \rightarrow 0$. As stated earlier, the resulting problem corresponds to a double edge notch test specimen of crack length $a$ and infinite width $b$. The shape correction factor for the energy release rate corresponding to this particular geometry can be obtained from \cite{Tada1973}, and is given by
\begin{linenomath}
\begin{equation}
	f \left( \dfrac{a}{b} \right) = \frac{\left[ 1.122 - 0.561 \left( \dfrac{a}{b} \right) - 0.205 \left( \dfrac{a}{b} \right)^2 + 0.471 \left( \dfrac{a}{b} \right)^3 -0.190 \left( \dfrac{a}{b} \right)^4 \right]^2}{1 - \dfrac{a}{b}}.
\end{equation}
\end{linenomath}
\noindent It is easy to see that for $b \rightarrow \infty$, the above reduces to $f \left( 0 \right) = 1.122^2 = 1.2589$. Note that this implies a strong discontinuity in $f_2 \left( 0, \lambda \right)$ at $\lambda = 1$, since \cite{Zak1963} have shown that for $E^\prime_1 < E^\prime_2$, the energy release rate must be zero when the crack tip is situated at the material interface. Furthermore, $f_2$ is non-monotone not only with respect to $\lambda$ in Case II-a, but also $\Upsilon$, as shown clearly in Figure \ref{fig:nonmonotonicity}.

\subsection{Remark on the accuracy of numerical simulations} \label{sec:accuracy}
\begin{figure}
	\centering
	\includegraphics[width=0.6\linewidth]{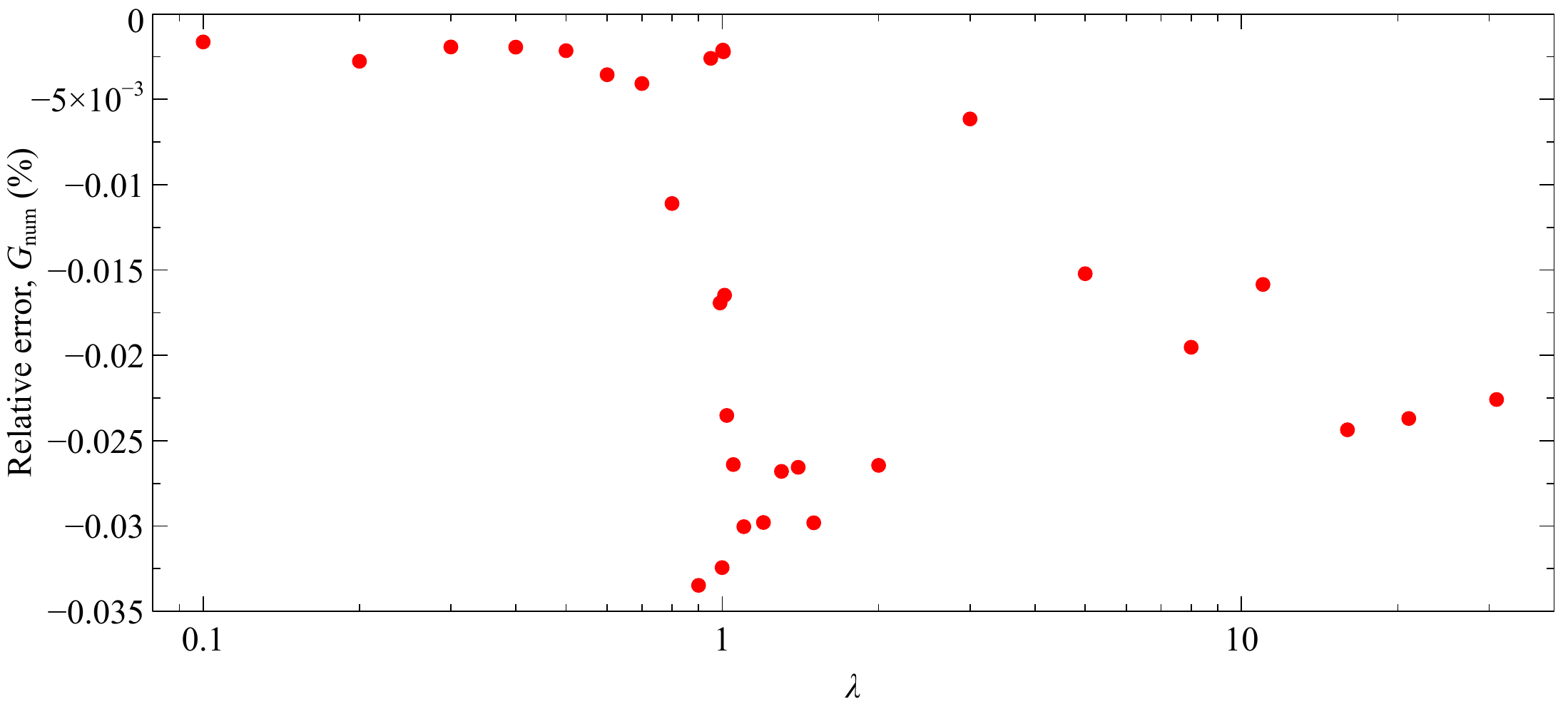}
	\caption{Relative errors of numerically calculated energy release rates with respect to the exact solutions for $E_1  = E_2$.}
	\label{fig:errorsInHomogeneousCase}
\end{figure}

The setup where in which $E_1 = E_2$ corresponds to the original problem studied by Griffith, for which the exact expression for the energy release is known. This allows us to determine the accuracy of our numerical results for this specific scenario. In particular, $\left( f_\text{num} - 1 \right) \cdot 100$ gives the relative error, in percent, of the numerically calculated energy release rates with respect to the exact solution. This is plotted in Figure \ref{fig:errorsInHomogeneousCase} for the range of crack lengths simulated in the current study.
We can observe that the errors are all negative. This is to be expected since displacement-based, fully compatible finite elements are known to give lower bound solutions with respect to the energy norm \cite[see for example,][] {Liu2008}. It should be noted that the meshes corresponding to the different prescribed crack lengths are independently generated, hence the plotted points are not inter-related and no trend can be inferred regarding the dependence of approximation errors on $\lambda$. On the other hand, the simulations yield results that are within $0.035\%$ of the exact solutions, even though actual crack lengths and the overall dimensions of simulation domains differ by several orders of magnitude.

\begin{figure}
	\centering
	\includegraphics[width=0.6\linewidth]{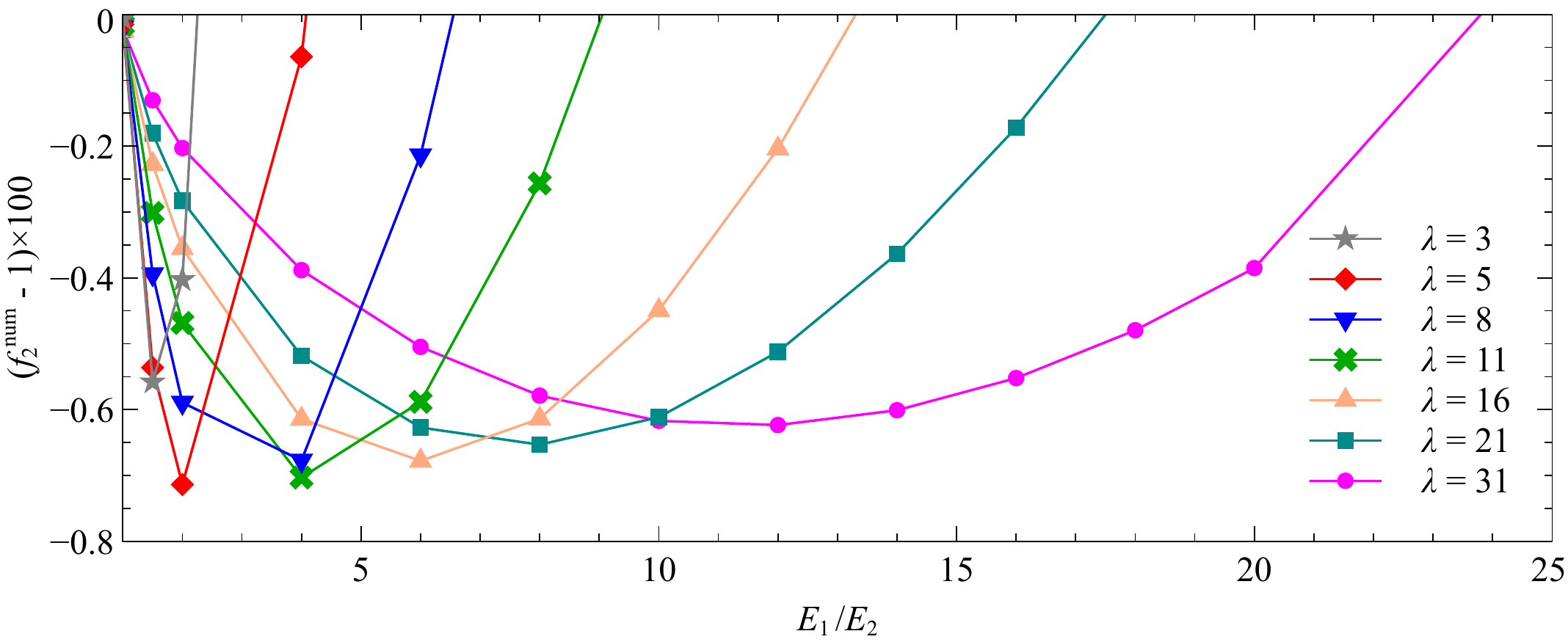}
	\caption{Percent relative errors with respect to unity of numerical results for correction factors $f_2 \left( \Upsilon, \lambda \right)$  where $\lambda \geq 1$ and $\Upsilon = E_1 / E_2 > 1$. The graph shows only the values associated with $\lambda$ and $\Upsilon$ for which $f_2$ is found to be less than 1.}
	\label{fig:errorsInHeterogeneousCase}
\end{figure}

When $E_1 \neq E_2$, the exact solutions are generally unknown, hence we can no longer directly calculate the actual errors associated with the FE solutions. Instead, we can exploit known characteristics of the energy release rate in some cases to infer some idea on the accuracy of our numerical results. Specifically in the case where $\lambda > 1$ and $\Upsilon = E_1/E_2 > 1$, $f_2 \left( \Upsilon, \lambda \right) > 1$ and should converge towards unity for $\lambda \gg 1$. However as classical FEM underestimates the energy as mentioned previously, it is possible to obtain numerical results for which $f_2 < 1$. We in fact observe this in our simulations for $\lambda \geq 3$, as shown in Figure \ref{fig:errorsInHeterogeneousCase}.
The exact solution for $f_2$ should be greater than 1, thus the quantity $\left| f_2^\text{num} - 1 \right| \cdot 100$ represents a \emph{lower bound} for the percent absolute relative error of the numerically approximated correction factors with respect to exact solutions. The largest such error bound is found to be 0.714\%, which occurs for $\lambda = 5$ and $\Upsilon = 2$. From this we can infer that the absolute relative error for the worst individual approximation of a correction factor is \emph{at least} 0.714\%, and is most likely even greater. Such insight is important in order to avoid parameter over-fitting in closed form expressions intended to approximate the numerical results.

\subsection{Approximation of interface correction factor $a < L$}
As the material contrast is defined differently depending on the crack tip location (i.e. $\Upsilon = E^\prime_2\,/E^\prime_1$ when $\lambda < 1$, and $\Upsilon = E^\prime_1\,/E^\prime_2$ when $\lambda > 1$), the interface corrections factors $f_1 \left( \Upsilon, \lambda \right)$ and $f_2 \left( \Upsilon, \lambda \right)$ from \eqref{eq:normalizedEnergyReleaseRate_1} and \eqref{eq:normalizedEnergyReleaseRate_2} are distinct functions that must be modeled separately. To arrive at an appropriate expression for $f_1 \left( \Upsilon, \lambda \right)$, we first take a look at the limit case $\Upsilon \rightarrow 0$, for which the problem reduces to one involving a center cracked test specimen of width $2L$.\footnote{The case involving a center cracked specimen of finite width has been dealt with extensively in the existing literature, however the prevailing convention was to express shape correction factors associated with stress intensity factors (SIF). Since the current work makes use of energy release rates in place of the SIF, we modify the various expressions proposed by prior authors accordingly.} \cite{Koiter1965} has shown that $f_1 \left( 0, \lambda \right)$ follows a limiting behavior as $\lambda$ approaches unity. That is,
\begin{linenomath}
\begin{equation}
	\lim\limits_{\lambda \rightarrow 1} \left[ \sqrt{f_1 \left( 0, \lambda \right)} \sqrt{ 1-\lambda} \right] = \frac{4}{\pi^2 - 4}.
	\label{eq:KoiterLimit}
\end{equation}
\end{linenomath}
A relatively simple closed form approximation for $f_1 \left( 0, \lambda \right)$ was proposed by \cite{Feddersen1967}, of the form
\begin{linenomath}
\begin{equation}
	f_1 \left( 0, \lambda \right) = \sec \frac{\pi \lambda }{2}.
\end{equation}
\end{linenomath}
This was later improved upon by \cite{Tada1973}, who added a polynomial factor in order to improve the accuracy of approximation. The resulting expression is given by
\begin{linenomath}
\begin{equation}
	f_1 \left( 0, \lambda \right) = \left( 1 - 0.025 \lambda^2 + 0.06 \lambda^4 \right)^2 \sec \frac{\pi \lambda }{2},
	\label{eq:TadaFormula}
\end{equation}
\end{linenomath}
which has a reported accuracy of 0.1\% for the original pertaining to stress intensity factor correction. In addition, the above formula reasonably satisfies the condition prescribed in \eqref{eq:KoiterLimit}.

\begin{figure}
	\centering
	\includegraphics[width=0.5\linewidth]{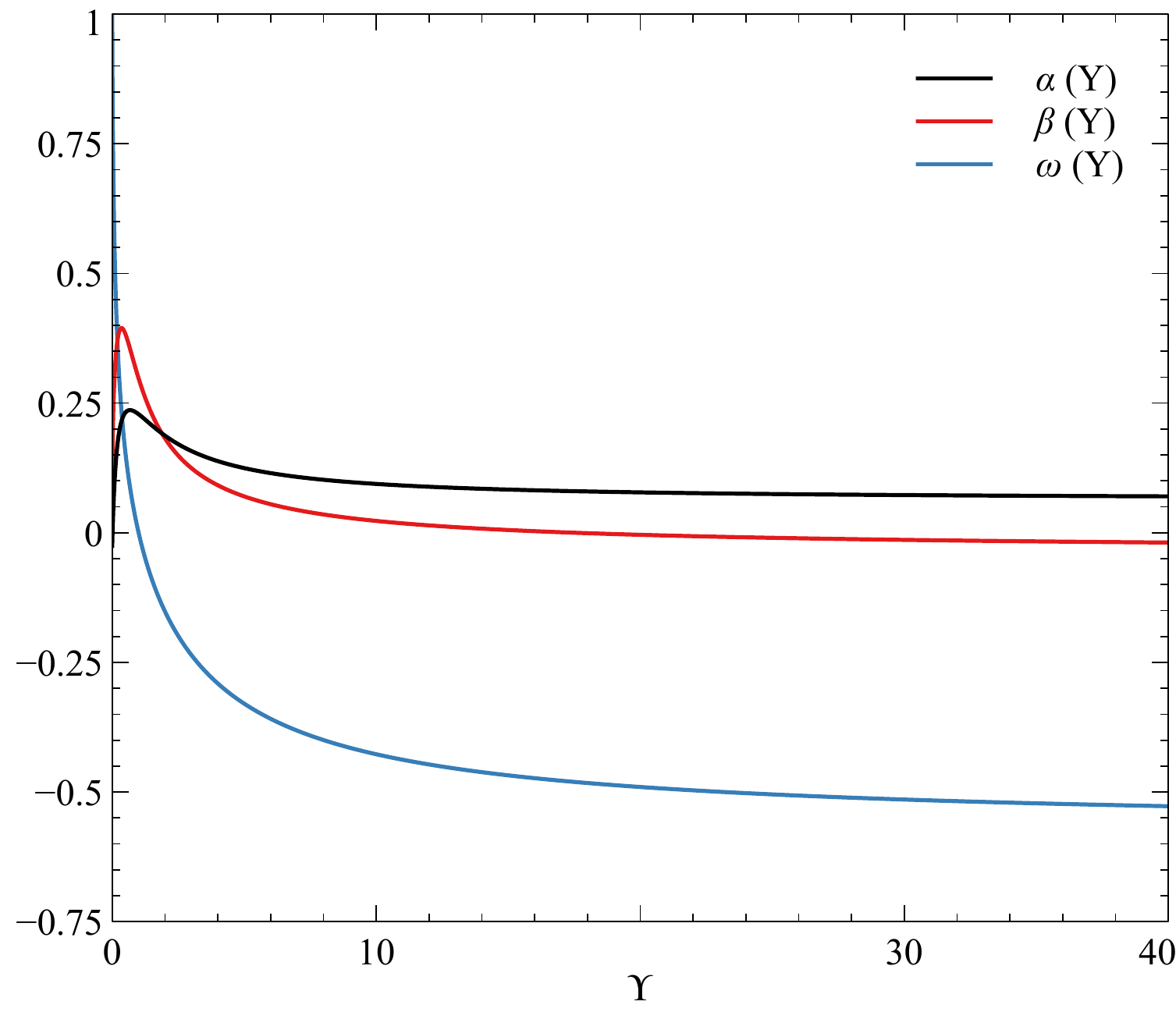}
	\caption{Behavior of functions $\alpha \left( \Upsilon \right)$, $\beta \left( \Upsilon \right)$ and $\omega \left( \Upsilon \right)$.}
	\label{fig:plot_f1_params}
\end{figure}

We propose a model for $f_1 \left( \Upsilon, \lambda \right)$ by generalizing the preceding formula to admit the material contrast as an additional argument. To do this, we replace the coefficients of the polynomial terms in \eqref{eq:TadaFormula} with functions of $\Upsilon$, and introduce an exponent $\omega \left( \Upsilon \right)$ that acts on both the polynomial factor and the secant term. This yields
\begin{linenomath}
\begin{equation}
	f_1 \left( \Upsilon, \lambda \right) = \left[ \left( 1 + \alpha \left( \Upsilon \right) \lambda^2 + \beta \left( \Upsilon \right) \lambda^4 \right)^2 \sec \frac{\pi \lambda}{2} \right]^{\omega \left( \Upsilon \right)},
	\label{eq:f1_model}
\end{equation}
\end{linenomath}
where now the main task is to obtain appropriate expressions for $\alpha$, $\beta$ and $\omega$. We have found it adequate to model these functions as a sum of different inverse power law terms plus a constant, i.e.
\begin{linenomath}
\begin{align}
	\alpha \left( \Upsilon \right) &= C_0^\alpha + \sum_{i=1}^3 C_i^\alpha \left( \Upsilon + B_i^\alpha \right)^{M_i^\alpha} 
	\label{eq:alpha} \\
	\beta \left( \Upsilon \right) &= C_0^\beta + \sum_{i=1}^3 C_i^\beta \left( \Upsilon + B_i^\beta \right)^{M_i^\beta}
	\label{eq:beta} \\
	\omega \left( \Upsilon \right) &= C_0^\omega + \sum_{i=1}^3 C_i^\omega \left( \Upsilon + B_i^\omega \right)^{M_i^\omega},
	\label{eq:omega}
\end{align} 
\end{linenomath}
wherein $M_i^\alpha, M_i^\beta, M_i^\omega < 0$. It is assumed that $\Upsilon \in \left[ 0, \infty \right)$, hence the proposed model should apply to both case I-a and I-b, and moreover is continuous and differentiable at $\Upsilon = 0$. The unknown parameters are determined by making use of a combined differential evolution and particle swarm (DEPS) optimization algorithm to fit the model given by \eqref{eq:f1_model}--\eqref{eq:omega} to the numerical results, with the objective that for each data point, the relative error between $f_1^\text{\,num}$ and the approximate model should lie within $\pm 1$\%.\footnote{Based on the findings in Section \ref{sec:accuracy}, imposition of a tighter tolerance would most likely result in over-fitting of the parameters and do little to improve the overall accuracy of the model. We note that while the adopted tolerance in fitting the model is much looser compared to that associated with \eqref{eq:TadaFormula}, the former is more than sufficient for applications involving subsurface fracture as the errors arising from uncertainty in material properties of different rock layers will most likely be the dominating factor affecting the accuracy of calculated results.} The constant terms are fitted first, utilizing the FE results for the limit case $\Upsilon \rightarrow \infty$. Here we found that $C_0^\omega$ converges very close to $1 - \pi/2$, thus we adopt said value for the exponent and fit $C_0^\alpha$ and $C_0^\beta$ accordingly. The remaining parameters are then optimized simultaneously, with the following constraints:
\begin{enumerate}[a)]
	\setlength{\itemindent}{10pt}
	\setlength{\itemsep}{0pt}
	\item $\alpha \left( 0 \right) = -0.025$, $\beta \left( 0 \right) = 0.06$ and $\omega \left( 0 \right) = 1$, in order to recover \eqref{eq:TadaFormula} in the limit $\Upsilon \rightarrow 0$,
	\item $\omega \left( 1 \right) = 0$, so that $f_1 \left( 1, \lambda \right) = 1$ when $E_1^\prime = E_2^\prime$.
\end{enumerate}
Parameter values arising from the optimization are summarized in Table \ref{tab:caseI_parameters}, with the resulting functions $\alpha \left( \Upsilon \right)$, $\beta \left( \Upsilon \right)$ and $\omega \left( \Upsilon \right)$ plotted in Figure \ref{fig:plot_f1_params}.
\begin{table}
	\centering
	\caption{Values of expression parameters for $\alpha \left( \Upsilon \right)$, $\beta \left( \Upsilon \right)$ and $\omega \left( \Upsilon \right)$.}
	\label{tab:caseI_parameters}
	\begin{tabular}{crrr}
		\toprule
		Parameter & $\alpha \left( \Upsilon \right)$ \hspace{2em} & $\beta \left( \Upsilon \right)$ \hspace{2em} & $\omega \left( \Upsilon \right)$ \hspace{2em} \\
		\midrule
		$C_0$ & 0.06432 28681 & $-$0.03714 40326 & $1 - \pi/2$ \hspace{1.4em}  \\
		\midrule
		$C_1$ & $-$0.00013 08622 & 0.45499 62001 & 0.00414 03509 \\
		$B_1$ & 0.15394 59365 & 0.16493 16577 & 0.01993 68747 \\
		$M_1$ & $-$2.97707 14455 & $-$0.86877 18544 & $-$1.02892 38200 \\
		\midrule
		$C_2$ & 0.72783 66450 & $-$0.00984 75618 & 1.49667 55116 \\
		$B_2$ & 1.43193 01449 & 0.03888 87677 & 2.05893 09171 \\
		$M_2$ & $-$1.28969 04253 & $-$1.14296 15885 & $-$0.94901 16025 \\
		\midrule
		$C_3$ & $-$0.15217 87410 & $-$0.41327 73403 & 0.06364 99752 \\
		$B_3$ & 0.53164 26030 & 0.69484 81483 & 0.20791 82322 \\
		$M_3$ & $-$1.92360 65137 & $-$3.84842 91265 & $-$1.41127 70181 \\
		\bottomrule
	\end{tabular}
\end{table}
Using said values, the calculated interface correction factor $f_1$ has a maximum absolute relative error of 0.84\% with respect to the FE solutions for $\Upsilon \neq 0$.

\subsection{Approximation of interface correction factor for $a > L$}
When $a > L$, the crack tip lies in the outer layer and we define the material contrast as $\Upsilon = E^\prime_1 / E^\prime_2$. As we can observe in Figures \ref{fig:fnum_2A} and \ref{fig:fnum_2B}, there is a fundamental difference in the behavior of $f_2 \left( \Upsilon, \lambda \right)$ between cases II-a ($\Upsilon < 1$) and II-b ($\Upsilon > 1$). Within the context of linear elastic fracture mechanics, a crack originating within the inner layer should not be able to cross the material interface due to the fact that $f \left( \Upsilon, 1 \right) = 0$ when $\Upsilon < 1$. Thus we ignore case II-a in the present work and construct an approximation for $f_2$ only in the case where $\Upsilon > 1$, for which we propose the expression
\begin{linenomath}
\begin{equation}
	f_2 \left( \Upsilon, \lambda \right) = \left\{ 1 + \gamma \left( \Upsilon \right) \left( \lambda - 1 \right)^\kappa e^{-\eta \left( \Upsilon \right) \left( \lambda - 1 \right)} \left[ 1 + \frac{1}{\sinh \left( \mu \left( \Upsilon \right) \left( \lambda - 1 \right) \right)} \right] \right\}^{\theta \left( \Upsilon \right)}.
\end{equation}
\end{linenomath}
\noindent Here, $\kappa = 0.19401\ 90296$ and the functions $\gamma \left( \Upsilon \right)$, $\eta \left( \Upsilon \right)$, $\mu \left( \Upsilon \right)$ and $\theta \left( \Upsilon \right)$ have the form
\begin{linenomath}
\begin{subequations}
	\begin{align}
		\gamma \left( \Upsilon \right) &= C_0^\gamma + \sum_{i=1}^3 C_i^\gamma \left( \Upsilon + B_i^\gamma \right)^{-M_i^\gamma} + C_4^\gamma \left( \ln \Upsilon \right)^{\,M_4^\gamma} \label{eq:gamma} \\
		\eta \left( \Upsilon \right) &= \sum_{i=1}^3 C_i^\eta \left( \Upsilon + B_i^\eta \right)^{-M_i^\eta} \label{eq:eta} \\
		\mu \left( \Upsilon \right) &= \sum_{i=1}^2 C_i^\mu \left( \Upsilon + B_i^\mu \right)^{-M_i^\mu} \label{eq:mu} \\
		\theta \left( \Upsilon \right) &= C_1^\theta \left[ \left( 1 + B_1^\theta \right)^{-M_1^\theta} - \left( \Upsilon + B_1^\theta \right)^{-M_1^\theta} \right] + C_2^\theta \left( \ln \Upsilon \right)^{\,M_2^\theta} \label{eq:theta}
	\end{align}
\end{subequations}
\end{linenomath}
The relevant parameters for the above expression are given in Table \ref{tab:caseIIa_parameters}, and result in $f_2 \left( \Upsilon, \lambda \right)$ having a maximum absolute relative error of 0.92\% with respect to the FE solutions where $1.001 \leq \lambda \leq 31$ and $1 \leq \Upsilon \leq 20$.
\begin{table}
	\centering
	\caption{Values of expression parameters for $\gamma \left( \Upsilon \right)$, $\eta \left( \Upsilon \right)$, $\mu \left( \Upsilon \right)$ and $\theta \left( \Upsilon \right)$.}
	\label{tab:caseIIa_parameters}
	\begin{tabular}{crrrr}
		\toprule
		Parameter & $\gamma \left( \Upsilon \right)$ \hspace{2em} & $\eta \left( \Upsilon \right)$ \hspace{2em} & $\mu \left( \Upsilon \right)$ \hspace{2em} & $\theta \left( \Upsilon \right)$ \hspace{2em} \\
		\midrule
		$C_0$ & 1.78387 28053 &&& \\ 
		\midrule
		$C_1$ & 0.40864 73188 & 2.32322 76308 & 3.05023 77162 & 1.25043 39256 \\
		$B_1$ & $-$0.93693 50334 & 9.67883 56763 & 0.97527 11044 & 1.16196 85473 \\
		$M_1$ & 0.07658 18295 & 0.72885 85092 & 0.41398 42077 & 0.40847 68000 \\
		\midrule
		$C_2$ & $-$2.17650 94908 & 7.24103 61595 & 10.10868 76917 & 0.03472 54619 \\
		$B_2$ & 6.28687 26947 & $-$0.99825 02642 & 0.93239 14535 & \\
		$M_2$ & 0.29679 87734 & 2.83648 89172 & 1.85287 26171 & 0.55275 40241 \\
		\midrule
		$C_3$ & $-$1.32027 18301 & 13.09075 19006 && \\
		$B_3$ & 10.33066 49702 & 0.98694 86003 && \\
		$M_3$ & 0.58213 56337 & 12.84741 14065 && \\
		\midrule
		$C_4$ & $-$0.64770 14404 &&& \\
		$M_4$ & 0.11464 72995 &&& \\
		\bottomrule
	\end{tabular}
\end{table}
Note that while we have conducted simulations involving larger material contrasts, the numerical results from such simulations where $\Upsilon > 20$ were not directly used in determining the aforementioned parameter values. Instead, we utilize them as a gauge of the model accuracy beyond the range of material contrasts used in the parameter fitting. For instance, the resulting maximum absolute relative error for $f_2$ when $\Upsilon = 50$ was found to be 3.57\%, which rises to 7.19\% when $\Upsilon = 100$.

\begin{figure}
	\centering
	\begin{subfigure}{0.45\linewidth}
		\includegraphics[width=\textwidth]{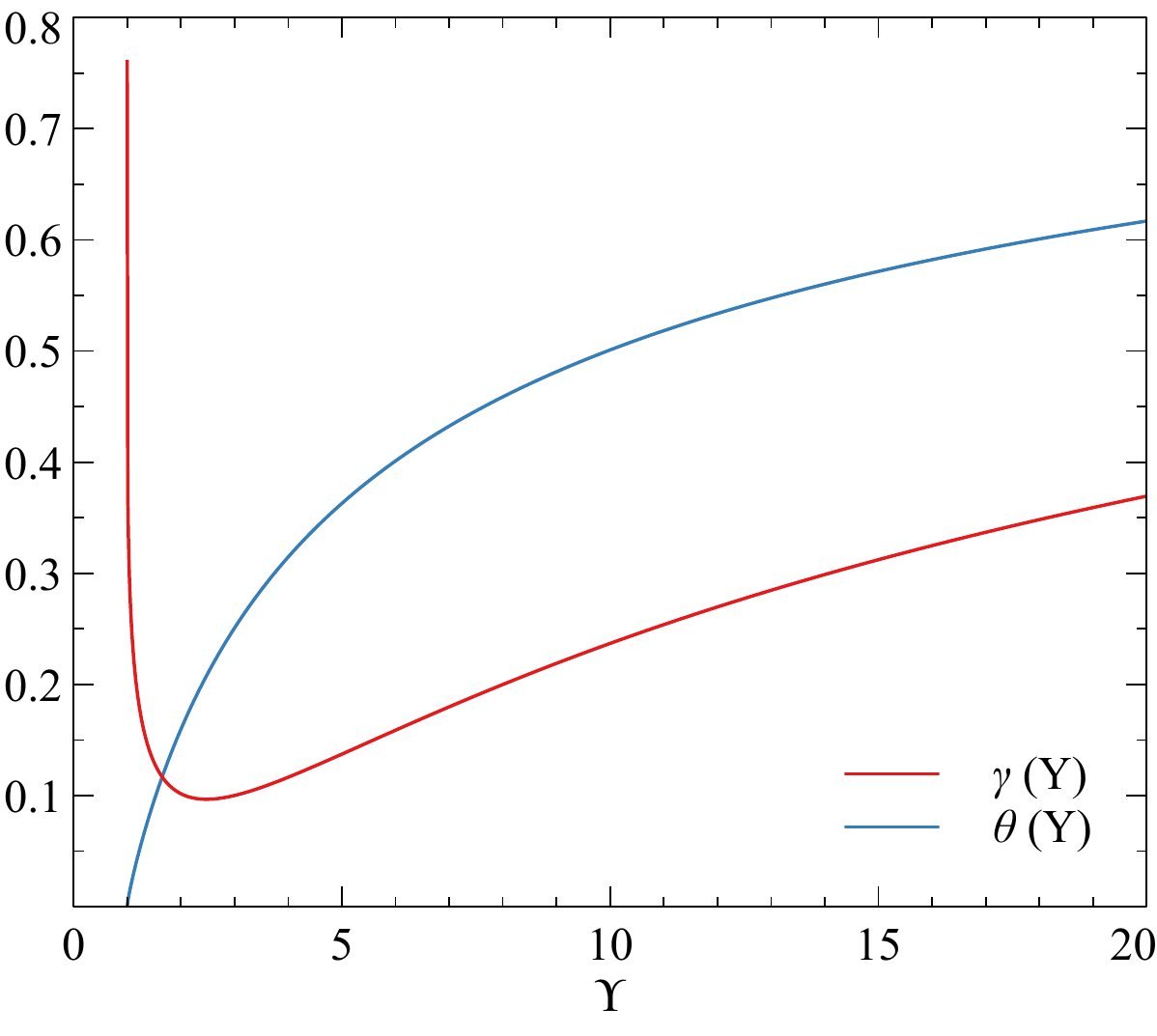}
		\caption{}
	\end{subfigure} \hspace{1cm}
	\begin{subfigure}{0.45\linewidth}
		\includegraphics[width=\textwidth]{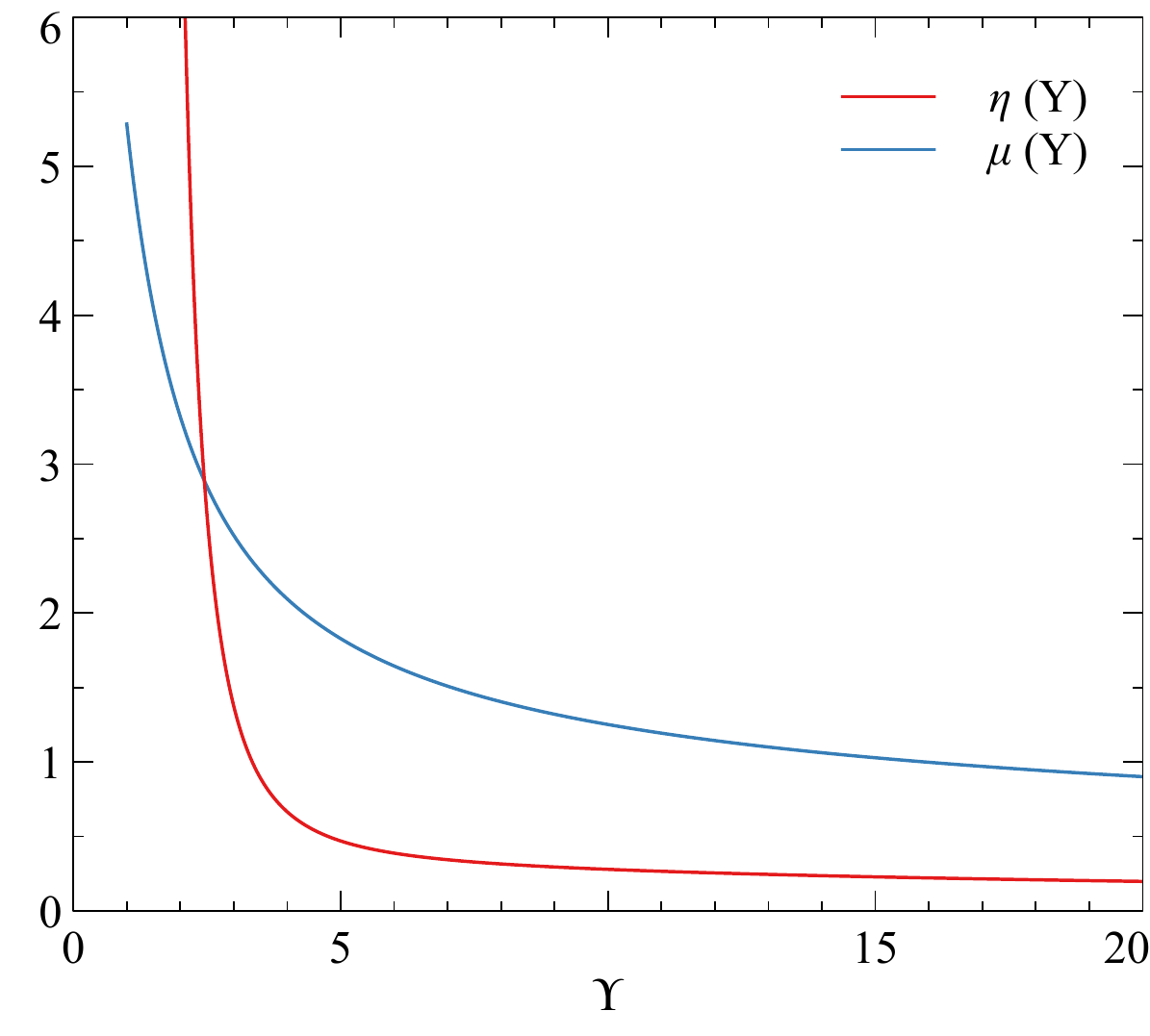}
		\caption{}
	\end{subfigure}
	\caption{Behavior of functions $\gamma \left( \Upsilon \right)$, $\eta \left( \Upsilon \right)$, $\mu \left( \Upsilon \right)$ and $\theta \left( \Upsilon \right)$ within the range of simulated material constrasts ($1 \leq \Upsilon \leq 20$) used for determining $f_2 \left( \Upsilon, \lambda \right)$.}
	\label{fig:plot_f2_params}
\end{figure}

The behavior of functions $\gamma$, $\eta$, $\mu$ and $\theta$ with respect to $\Upsilon$ are shown in Figure \ref{fig:plot_f2_params}.
We can see from \eqref{eq:theta} that $\theta \left( 1 \right) = 0$, which is necessary in order to achieve $f_2 \left( 1, \lambda \right) = 1$. Furthermore, constraints are imposed when fitting the parameters such that $B_i \geq -1$ in all the relevant terms of the parameter expressions, so that $\gamma$ and $\eta$ and remain smooth and bounded for $\Upsilon \geq 1$. This avoids the scenario where $f_2$ is indeterminate (specifically, $\infty^0$) when $\Upsilon = 1$. On the other hand, $\gamma$ and $\theta$ grow unbounded while $\eta$ and $\mu$ go to zero as $\Upsilon$ goes to $\infty$. Together, these result in the behavior where $f_2 \left( \Upsilon, \lambda \right) \rightarrow \infty$ as $\Upsilon \rightarrow \infty$, which is in line with the hypothesis that a limiting case does not exist for $f_2$ corresponding to an infinite $\Upsilon$.

%% file: concluding_remarks.tex
\section{Concluding remarks}
In the present paper, the energy release rate for a mode-I crack perpendicular to a material interface in a symmetric 3-layer configuration was investigated by means of finite element simulations. To enable the generalization of results, key quantities were recast into dimensionless groups, with focus placed on the combined effect of the crack length and the difference in material properties of the inner and outer layers. We found that it is possible to express the normalized energy release rate as the product of a simple linear relation and a correction factor accounting for proximity of the crack tip to a material interface. We assumed that the latter is a function purely of the normalized crack length and the material contrast, and developed closed form expressions that yield energy release rate predictions to within one percent of FE results.

One important application of the formulas developed in the current work is modelling the growth of fractures in mechanically layered geological successions. These formulas can allow us to predict whether fractures will remain confined within specific mechanical layers, forming layer-bound fracture networks \citep[as described in][]{Welch2019}, or will propagate across the mechanical layer boundaries forming through-cutting fractures. This will depend not only on the layer thickness and the contrast in mechanical properties between the layers, but also on the in situ stress and the duration of deformation. Using subcritical fracture propagation theory, we can calculate the propagation velocity of a fracture tip from the energy release rate. We can thus use the results described here to determine the conditions under which a fracture can propagate across a mechanical boundary within a realistic geological time due to stress corrosion. This work will be described in a forthcoming paper. We also plan to generalise the formulas to model fracture propagation across multiple mechanical boundaries, so that we can predict likely fracture connectivity in geological successions comprising many mechanical layers. This will have important applications for modelling fluid flow in the subsurface.

An unfortunate drawback of LEFM is its inability to allow for crack propagating from a compliant layer into a stiff layer, irregardless of the latter's thickness, due to the energy release rate being theoretically zero when the crack tip lies exactly on the material interface. Such a constraint is unrealistic compared to what can be observed for example in outcrops, and is due to the fact that LEFM does not take into consideration the material strength, but only relies on the energy release rate. This can be remedied by introducing alternative fracture theories that simultaneously take into account both the material strength and fracture energy. For example, \cite{Adams2015} adopted a cohesive zone model to determine critical values of the generalized stress intensity factor for a perpendicular crack impinging on a material interface. Another alternative for brittle materials is Finite Fracture Mechanics \citep[see for instance][]{Weissgraeber2016}. It should be noted that once the crack tip propagates past the interface, the crack tip stress distribution recovers the inverse square root singularity, so  that evolution theories based on the energy release rate can again be used to model the crack extension over time. However this would also entail having to construct the approximate expressions for the interface correction factor associated with Case II-a in the previous discussions. Furthermore, crack advance under weakly singular crack tip stresses results in an instantaneous finite extension of the fracture, implying a jump discontinuity of the crack tip location over time. For the moment, both are outside the scope of the current paper and will be addressed in future work.